# On Constant Distance Spacing Policies for Cooperative Adaptive Cruise Control

Kay Massow, Ilja Radusch, and Robert Shorten

*Abstract*— **Cooperative Adaptive Cruise Control (CACC) systems are considered as key potential enablers to improve driving safety and traffic efficiency. They allow for automated vehicle following using wireless communication in addition to onboard sensors. To achieve string stability in CACC platoons, constant time gap (CTG) spacing policies have prevailed in research; namely, vehicle interspacing grows with the speed. While constant distance gap (CDG) spacing policies provide superior potential to increase traffic capacity than CTG, a major drawback is a smaller safety margin at high velocities and string stability cannot be achieved using a one-vehicle look-ahead communication. The hypothesis of this work is to apply CDG only in few driving situations, when traffic throughput is of highest importance and safety requirements can be met due to comparably low velocities. As the most relevant situations where CDG could be applied, we identify starting platoons at signalized intersections. In this paper, we illustrate this idea. Specifically, we compare CTG with CDG regarding its potential to increase the capacity of traffic lights. Starting with the elementary situation of single traffic lights we expand our scope to whole traffic networks including several thousand vehicles in simulation. Using real world data to calibrate and validate vehicle dynamics simulation and traffic simulation, the study discusses the most relevant working parameters of CDG, CTG, and the traffic system in which both are applied.**

## I. Introduction

CACC is the extension of Adaptive Cruise Control (ACC), a driver assistance system which automatically adjusts the speed of a road vehicle to maintain a safe distance from vehicles ahead [1]. Today's ACC systems use radar sensors to measure this distance. CACC extends ACC by additional communication components to exchange information with preceding vehicles. This information exchange helps to increase the density of platoons of vehicles with activated ACC and to potentially tackle string instabilities occurring in such platoons. String instability in vehicle platoons is caused by radar sensor delays and the dynamics of the vehicles and their power trains. To achieve overall string stability, constant time gap (CTG) spacing policies have prevailed in research, i.e. the target distance between vehicles grows with the speed. However, growing distances entail efficiency loss. This fact is reflected by the recent decision of Daimler to cancel their truck platooning program, which aimed on a 0.68 seconds time gap (15 m at 80 Km/h) and did not achieve the expected efficiency in terms of fuel saving [2].

In this work, a constant distance gap (CDG) policy for CACC is considered. Although CDG can improve traffic throughput enormously, its applicability in real traffic has been proven to be very limited, as it is not suitable to achieve robust string-stability without significant loss of efficiencies [3]. The hypothesis of this work is to apply CDG only in few driving situations when the following circumstances occur:

- traffic throughput is of high importance.
- platoon sizes are short enough that string instability or communication topology complexity can be handled, e.g. employing mini-platoon control strategy [3].
- velocities are low enough to cover safety requirements, acceleration is smooth and predictable.

While there are several use cases in which such conditions prevail, clearly, traffic-light-controlled intersections are one of the most relevant. In particular, the traffic flow of two crossing streets share one spot in a time duplex manner. Thus, exhibiting the highest possible traffic density on this spot is of high importance. Intersections controlled by traffic lights in addition provide clearly regulated right-of-way, i.e. during a green light phase, a platoon can pass this spot as a whole without paying attention to the cross traffic. Moreover, starting up from a stop line when the traffic light changes to green results in a smooth and predictable acceleration maneuver. Thus, we will focus on traffic-light-controlled intersections, with other situations presented in future work. In what follows we shall assume urban speeds of up to 50 km/h and stable platoons on intersection either achieved by limited length or capable communication topologies [3]. The research questions discussed in the rest of this paper focus on capacity improvement of CDG over CTG at signalized intersections. Our model for car following dynamics is based on the controller design presented in [31], parameterized using real world data.

K. Massow is with the Daimler Center for Automotive IT Innovations at Technical University Berlin, Berlin, Germany (e-mail: kay.massow@dcaiti.com).

I. Radusch is with Fraunhofer Institute FOKUS Berlin, Berlin, Germany (e-mail: ilja.radusch@fokus.fraunhofer.de).

R. Shorten is with the Dyson School of Design Engineering, Imperial College London, SW7 2AZ London, U.K. (e-mail: r.shorten@imperial.ac.uk).

**Comment: Before proceeding some comments on string stability are in order.** *Although string stability is an important aspect for realizing CDG in platoons (see related work in the next section), we do not address string stability nor related control theory in this work. Instead we focus on assessing the traffic performance of CDG over other spacing policies. While there are many other publications dealing with string stability, the rationale for this work is the usefulness of platoons, string stability permitting, in the context of specific use-cases. Our objective here is to study one such situation in detail, and to illustrate the effectiveness of platoons in an elementary situation in which string stability is not likely to be a serious technical issue.*

### A. Main findings of this work and the structure this paper

Assessing the benefit of CDG for signalized intersections requires a comprehensive and thorough consideration of a whole traffic system. This includes many microscopic and macroscopic aspects and aggregating partial results. From the authors' perspective, these should be presented as a whole and not be split apart in different papers. With this in mind our paper is structured as follows. After discussing related work in the Section II, the remainder of the paper is structured as follows.

- In Section III, we define our research scope and asses the CDG capacity improvement at a single traffic light on a straight road. For this purpose, we parameterize a CDG policy for vehicle simulation using real world data. CDG shows traffic throughput improvement over the CTG baseline of up to 140%.
- In Section IV, we extend our study to a whole intersection, in order to cover traffic related aspects which lower the throughput, e.g. turning vehicles and right-of-way. Vehicle simulations, including 160 vehicles, showed that these aspects can lower the CDG throughput improvement down to 27% in worst case. We further found that CDG benefit on throughput grows superlinear with the CDG penetration rate among vehicles.
- In Section V, we present a method to calibrate a traffic simulation model using vehicle dynamics simulation. This is a prerequisite to employ many thousands of vehicles to simulate CDG in a whole traffic system, while generating the same results as the vehicle dynamics simulation regarding relevant metrics.
- In Section VI we study the impact of CDG on mutually influencing intersections of a traffic system. A synthetic arterial scenario of five intersections revealed that CDG situationally creates backlogs of adjacent intersections, which block the cross traffic. A synthetic grid scenario of 25 intersections revealed that CDG is vulnerable to create gridlocks. We show the impact of these effects on traffic throughput and how to tackle them by adjusting the traffic light configurations regarding green light times and offset.
- In Section VII, we complement our findings with studying CDG in a real world road network simulation scenario including ten intersections in Berlin, Germany. CDG gains a throughput improvement of 70%, while a penetration of 50% CDG reached an improvement of 25%. To exhibit its full potential in urban traffic, CDG needs to enable cutting in and to prevent junction blocking by cooperative aspects.

We conclude this paper in Section VIII. In order to help the reader to follow the findings arising throughout the study, each section concludes with a discussion of its main findings.

## II. RELATED WORK

The most relevant goals for the design of CACC systems are to create small gaps between vehicles to increase road capacity, guarantee string stability [5], while keeping the communication topology realizable, i.e. as simple as possible [1]. The latter is, in the best case, reduced to each vehicle in a platoon receiving data from its direct preceding vehicle. Further possible communication structures may include receiving data from the platoon leader, multiple predecessors, the successor, or from a fully networked platoon [26]. Each of them entail different advantages regarding control quality, string stability and, thus, on the minimum gap size. Further goals on control optimization are ride comfort and fuel/energy consumption, which are both dependent from acceleration profiles.

### A. Constant Time Headway Policy (CTG)

The constant time gap policy refers to maintaining a time gap between vehicles in a platoon, which means that the gap increases with the velocity. It has received most attention in the literature for being known to improve string stability even with the simplest communication structure [5], [6]. It also contributes to safety, driving comfort, and imitates the human driver behavior. However, the downside of the velocity dependent gaps is the platoon length growing with the velocity and the associated required road space. Even very small time-gaps of 0.6 s [7] relates to additional road space of 8 m at 50km/h compared to stand still.

### B. Constant Distance Gap Policy (CDG)

The constant distance gap policy refers to a fixed gap between vehicles, independent form the velocity. This policy entails the maximum efficiency in terms of road capacity, however string stability cannot be achieved using the information of the preceding vehicle only. In [8] it was shown that including additional information from the platoon leader is required. In order to address string stability, further communication topologies like mini-platoons or multiple vehicles look ahead are reviewed in [3]. Cyclic as well as bidirectional communication is discussed in [9].

### C. Adaptive Headway Policies

There are different approaches that either combine CDG and CTG in one policy, or further include different control goals by making the gap dependent from more parameters than velocity. In [10] a variable time gap (VTG) policy is proposed, taking traffic flow aspects into account for calculating the desired gap. Further work has been done to integrate safety aspects and vehicle limitations in the spacing [11], or to adapt it to human behavior [12]. These adaptive policies gain their positive effect mostly at shorter distances at lower speeds compared to CTG.

### D. Cooperative Maneuvers Regarding Cross/Parallel Traffic

Another important aspect regarding the spacing of CACC platoons, is related to cooperative maneuvering [13]. Since platoons need to allow for cut-in maneuvers of other vehicles, the required gaps have to be provided on demand. For urban applications, cooperation is especially required at intersections when platoons need to be crossed by other vehicles. Such applications [14] which extend CACC to accommodate cross traffic and parallel traffic are currently researched e.g. in the German research project IMAGinE [37]. Its applications relevant for remainder of this work are cooperative lane merging and cooperative decentralized intersection to ensure clearing intersections for cross traffic.

### E. Cooperative Start-Up at Intersections

In the field of combining CACC with traffic-light control, most research aimed at synchronization of platoons and green lights phases, so that stop and go can be prevented, such as [15]. Very few works focus on start-up control coordinated among vehicles and traffic lights, so that as many vehicles as possible can pass an intersection after stand still. [16] studies platoons of vehicles waiting in front of a traffic-light regulated intersection, using SUMO [38]. A coordinated start-up initiated by a V2X message SPAT (SAE 2735) of the traffic light is proposed and the underlying algorithm also addresses the problem of low market penetrations. [17] considers a cooperative start-up of real world platoons at traffic lights. Findings indicate that a constant and preferably small gap is essential for increasing the throughput at traffic light regulated intersections. [18] presents an automatic start-up control to start up vehicles with less delay (47.2%) to improve traffic throughput, while [19] addresses an optimized acceleration profile to reduce fuel consumption.

### F. Platoons in Signalized Networks

In order to assess the impact of CACC on whole traffic systems, it is not sufficient to consider isolated traffic lights and intersections. In fact, multiple mutually influencing intersections such as signalized arterials need to be considered. This becomes especially relevant for dense platoons of vehicles passing. [20] presents and algorithm to optimize signals at arterials based on real-time platoon information. Different penetration rates are evaluated on an eight-intersection arterial using the VISSIM simulator. While most other research in this field focus on the control of traffic lights, [21] addresses optimization from the perspective of the vehicles in a cooperative way. Clusters of vehicles are formed that approach and depart at intersection on signalized arterials. The approach requires a penetration rate of 100% and aims on increasing traffic throughput while reducing energy consumption. [25] showed, by means of a 16-intersection arterial, that throughput can be doubled by reducing human delay and time gap without changing the signal control. [22] and [23] aim on preventing stops by slowing down until the queue waiting at the intersection starts moving. Penetration rates lower than 100% are considered in [22]. In [24] splitting up platoons and predicting trajectories aim on ideally passing green light phases. However, this requires a certain space while approaching the intersection and may hardly work for arterials with small intersection interspaces.

## III. SINGLE TRAFFIC LIGHT PERFORMANCE

In this section we begin researching the performance of CDG on a single traffic light, before considering whole intersections and traffic systems in the subsequent sections. For this purpose, we first need to define a baseline for comparison with other spacing policies and how performance can be measured.

In this regard, we define the research scope of this work, including preliminary assumptions. From this scope, we derive the determining working parameters for all policies e.g. the standstill distance, as these parameters have a big influence on the performance. Once these parameters are identified, we use real world data to calibrate them. Finally, we describe the implementation of the policies we use for simulation with the PHABMACS simulator [15] and we evaluate the results.

### A. Research Scope

The most relevant metric to assess traffic light performance is its capacity, which is defined by its maximum throughput, i.e. the maximum possible number of vehicles passing per time unit [27]. The relevant relationship between throughput and platoons passing the traffic-light, hence, is the number of vehicles per platoon length.

The portion of platoon length pertaining to each vehicle in a CTG platoon is depending on the parameters depicted in Fig. 1. The constant portion is the vehicle length plus the standstill distance, while the dynamic portion is the time gap, which grows with the platoon velocity. The dynamic part is zero in CDG platoons, i.e. the CDG platoon length is always the same like in standstill, which makes the CDG so effective.

Another relevant parameter, especially for the start-up at traffic lights, is the drivers' reaction time. This time refers to the delayed start-up of a vehicle in the platoon with regard to the start-up of its preceding vehicle. In contrast to CTG, which is similar to human drivers' vehicle following behavior, CDG can hardly be realized by humans. Thus, for CDG we assume a fully automated longitudinal control with no driver in the loop. This consideration is especially relevant for the start-up at traffic lights, as human reaction time would make notable difference here. Since we aim on comparing the following behavior of CDG with other policies, we neglect the reaction time for all policies in this work.

Accordingly, in order to compare CDG with CTG, we need to parametrize the constant portion, vehicle length and the stand still distance with the same values. Furthermore, these values should be chosen as realistic as possible for comparison, as their ratio to the time gap makes a considerable difference. Finally,

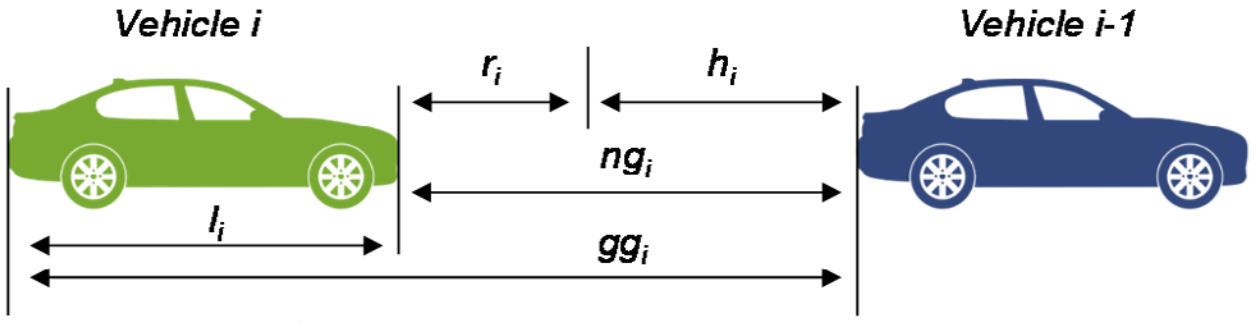

Fig. 1. A platoon of i vehicles, where $l_i$ is the length, $r_i$ is the standstill distance, $h_i$ is the time gap, $ng_i$ is the net gap, and $gg_i$ is the gross gap of the $i^{th}$ vehicle in the platoon

we also need to parameterize the time gap of CTG as realistic as possible.

Indications for all these parameters could be derived from HCM [27] and the German HBS [28]. The HCM indicates a capacity on average roads of 2400 vehicles per hour, while the HBS indicates 2000 vehicles per hour. Besides the fact that both values differ considerably (gross time between vehicles of 1.8 s and 1.5 s) we have no indication on how to split that time in the dynamic and the constant portion. Recent work [14] on the other hand indicates that time gaps for CTG of below 0.6 s (the dynamic portion only) can be realized for string stable platoons with automated CACC.

**Remark:** *Due to this large range of reasonable values, we decided not to define our baseline for comparison based on theoretic values from the standard works such as [27], nor on best possible time gaps achieved in current research, such as [14]. Instead we decide to assume for this study, that future CACC distance behavior in series production will be of similar performance as skilled human drivers and with no reaction time. For this purpose, we derive our baseline (time gap and standstill distance) from real world data collected during the field trial simTD [29]. For the sake of fairness, in this section, we will also present results of using parameterization of best achievable time gaps of current research. We further assume that the velocities in our study is low enough so that CDG can keep the standstill distance.*

The resulting parametrization is presented in the next subsection. Furthermore, as earlier mentioned, CDG should not be applied at arbitrary high velocities due to safety aspects and stability issues arising when the one-vehicle look-ahead communication pattern is applied. Thus, there is a speed limit at which the CDG spacing policy is required to be switched to CTG. As most traffic light scenarios are located in urban areas we limit our study to velocities below 50 Km/h. For the sake of completeness, we define and study a policy that switches from CDG to CTG at 30 Km/h. This policy will be referred to as SWITCH in the remainder of this work.

### B. *Calibration of Simulation on Real World Data*

As motivated in the previous subsection, we employ real world data to calibrate the policy parameters for simulation, as well as the baseline for evaluation. The data we used has been captured at simTD [29], a large scale field trial for testing V2X applications conducted over a period of six month, including a test fleet of 100 controlled vehicles. For the calibration of the simulation model, we consider start-up situations at traffic lights. The relevant calibration data for parameterization includes the acceleration profile in order to model the first vehicle of a platoon, the standstill distance and the time gap. Therefore, we filtered situations from the logged test data according to the following constraints:

- start-up after standstill, preceding vehicle is present;
- vehicle accelerates, target speed 40km/h – 65 km/h;
- accelerator is not released during the situation.

The filtered data included 3,546 start-up situations from 27,642 logged trips driven by 98 different drivers (73 male, 25 female). Fig. 2. depicts the resulting data, inspired by the model matching process for acceleration maneuvers described in [13]. All situations were aligned time-wise, at the point time when the preceding vehicle starts moving. The resulting curves of velocity and distance to the preceding vehicle were averaged. The averaged time gap settles at 0.87 s and the average standstill distance is 2.95m. We used these values to feed our simulation models. The black dotted lines represent the 95% confidence interval, which mark the band for simulation model validity according to [13]. We calibrated the acceleration profile of the platoon leader in our simulation to match the average speed trajectory of real world data. The speed profile in simulation matches the confidence band of the real world data, except for some dents in the graph during gear shifts. Thus, we consider the simulation model as valid representation of the real world data. In this way we were able to determine all relevant parameters as defined for our research scope, except for the vehicle length. For the vehicle length we assume 5.15 m due to the following considerations. According to [30] in 2011 we can assume an average length of passenger vehicles of 4.75 m. We add further 0.4 m to represent the increased length of vehicles since 2011 and some heavy duty traffic.

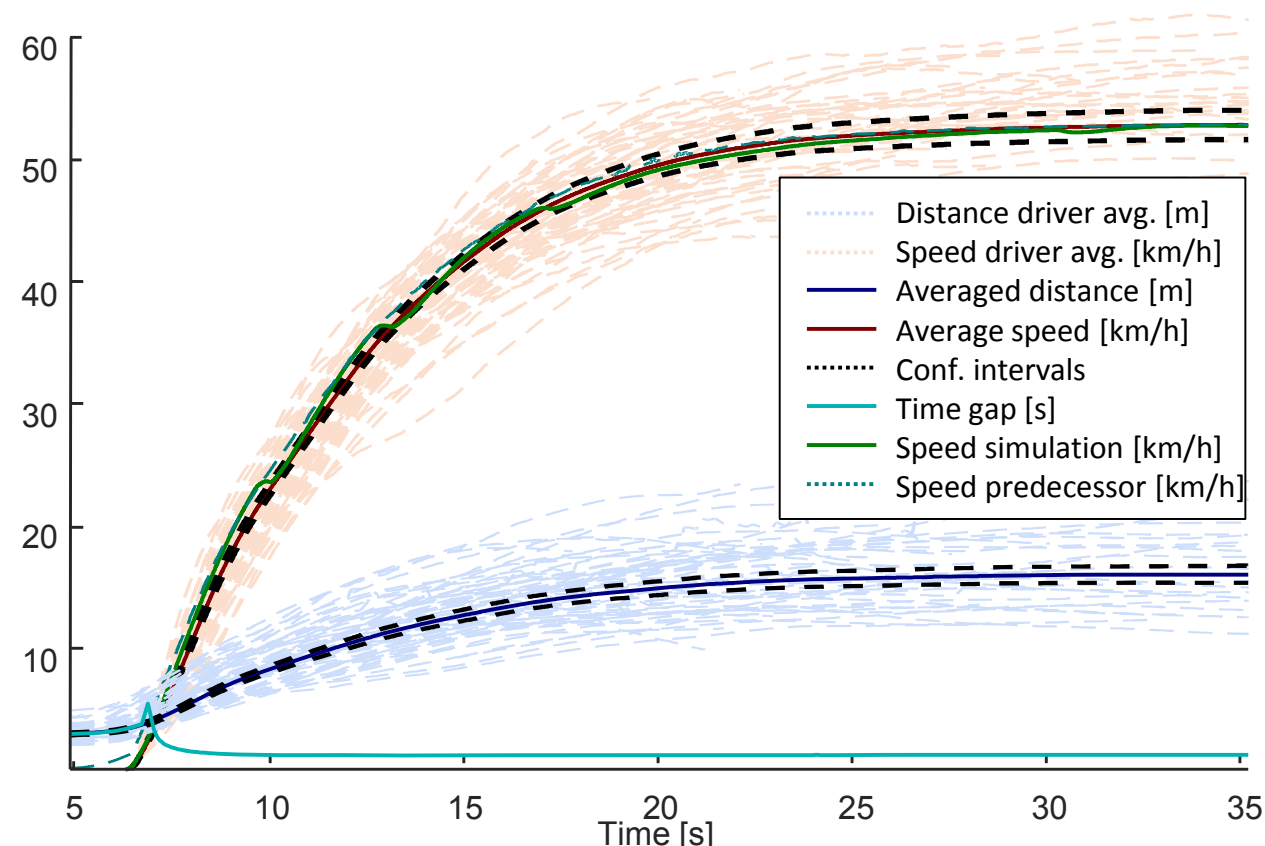

Fig. 2. Averaged Velocity and following distance processed from real world start-up situations to derive CDG and CTH controller parameters

### C. *Spacing Policies*

Using the parameters derived in the previous subsection, we finally define the following spacing policies for studies in simulation.

#### 1) *CDG-Constant Distance Gap*

The constant distance gap policy *CDG* is defined by the vehicle length of 5.15 m and the stand still distance of 2.95 m determined in the previous subsection.

#### 2) *CTG-Constant Time Headway*

According to the calibration with real world data we define the baseline policy for this work with 0.87 s time gap, in the following referred to as *CTG-Ref.* At 50 Km/h a gap of 15 m is reached. For comparison, we also define a policy *CTG-HCM* to match the American HCM at maximum speed in urban areas (50 Km/h). Assuming the gross time gap between vehicles of 1.5 s (HCM at 2400 vehicles per hour) together with vehicle length and standstill distance (as defined above), this results in a net time gap of 0.92 s. *CTG-HBS* represents the German HBS with 2000 vehicles per hour and, thus, with 1.22 s time gap.

*3) Switch*

Based on the parameters of *CDG* and *CTG-Ref* we define two polices to switch between both of them at a predefined velocity of 30 Km/h. *SWITCH-1* renders the time gap using the difference between the current velocity and 30 Km/h, i.e. at 50 Km/h a gap of 6 m is reached. *SWITCH-2* increases the gap from 0 m at 30 Km/h to 15 m at 50 Km/h, so that the same distance as with *CTG-Ref* is reached.

*4) Mix*

In order to enable studying a certain rate of CDG penetration, we define the *Mix* policy. The penetration rate is set to 50 % with a randomly alternating pattern on *CDG* and *CTG-Ref*.

### *D. Realization*

All spacing policies described above haven been implemented in the PHABMACS simulator [13], for subsequent evaluation. Further, all policies rely on the one-vehicle look-ahead communication pattern [26]. The evaluation scenario simply consists of a straight single lane road with a single traffic light, generated manually. In order to measure the maximum achievable throughput of all policies we create the same initial condition for each policy simulated. All vehicles approach the traffic light while red and queue up to standstill at the stop line. Once all vehicles have stopped the traffic light turns green and the platoon starts accelerating up to 50 Km/h. Vehicles passing the stop line are counted for evaluation.

*1) CTG*

The basis controller for the vehicles is a Java implementation of the cascaded PID framework presented in [31] (see Fig. 3), integrated as longitudinal controller in the PHABMACS driving controller hierarchy (see [13] for explanation). As the controller design is discussed in detail in [31], we just briefly describe its main components. $G_i$ represents the low-level controller $LL$ acting on the vehicle model $i$, where $i$ represents the $i^{th}$ vehicle in the platoon. $LL$ is different from the low-level controller in [5] and was initially presented in [32]. The input of $LL$ is the control value $u_i$ represented by the desired acceleration of the vehicle, while the output is the desired torque for the engine and the brake, which are fed directly to the vehicle model as described in [13]. $C_{i,ACC}$ is a PD-type feedback controller that acts on a locally sensed distance to the preceding vehicle with a simulated sensor delay of 150 ms. $H_i$ implements the spacing policy. For CTG the policy $H_i$ is defined by $1 + h_{d,i}s$ [31] (here s is the Laplace transform variable) which is the transfer function representation of $d_{r,i} = r_i + h_{d,i}v_i$ in the time domain, where $d_{r,i}$ is the desired spacing, $r_i$ is the standstill distance, $h_{d,i}$ is the time gap and $v_i$ the velocity. $C_{i,CACC}$ is a feedforward filter described in [31] using the communicated information of the directly preceding vehicle, i.e. the current and desired acceleration $a_{i-1}$ and $u_{i-1}$, as well as the current time lag of the vehicle model $\tau_{i-1}$. In contrast to [31] we treat $\tau_i$ as a dynamic value for each vehicle, which is taken online from a calibrated map depending on the current gear, requested torque (drive/brake), and current engine rotational speed.

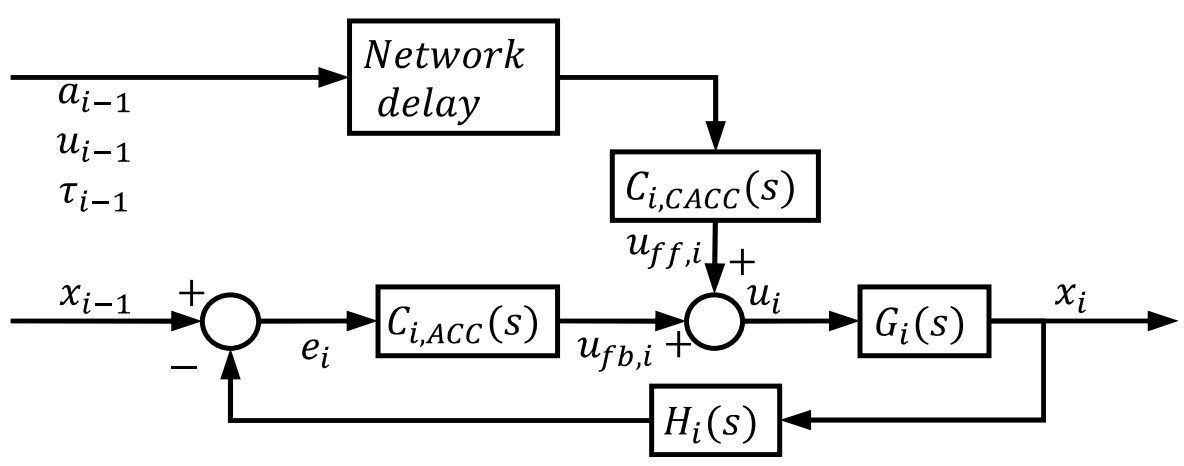

Fig. 3. Control structure of the longitudinal model

*2) CDG*

For the *CDG* policy, two aspects differ from the setup described above. The spacing policy $H_i$ is expressed by $d_{r,i} = r_i$ in the time domain and $H_i = 1$ in the frequency domain. The feedforward controller $C_{i,CACC}$ is the implementation of the optimization problem depicted in Fig. 4. An acceleration curve is predicted for the preceding vehicle, based on the received information $a_{i-1}(t)$, $u_{i-1}(t)$, and $\tau_{i-1}(t)$. Taking the latest measured communication delay into account, $u_i$ is calculated so that $a_i$ meets $a_{i-1}$ in a predefined time interval in the future.

*3) SWITCH*

By combining *CDG* and *CTG* according to the parameters described above, we realized *SWITCH* as a simple change between both policies at 30 Km/h.

### *E. Evaluation*

Fig. 5 depicts the results of six simulation runs with one graph each for the seven described polices. The graphs can be interpreted as a vehicle counter passing the traffic light stop line over time. The counter starts at time 0 when the traffic light turns green after red. The vertical lines in the figure mark the throughput of different green phase lengths. The throughput of the alternative spacing policies for a specific green phase length can be read from the figure at the point where its vehicle counter graph crosses the vertical lines. For instance, the number of passing vehicles at a green phase of 15 s is 17 for *CDG*, 13 for *SWITCH1/2*, 11 for *MIX*, 8 for *CTG-Ref*, and 7 for *CTG-HBS*. In order to enable comparing the performances of all policies at each phase, we exceptionally applied a deterministic alternating pattern on the *Mix* policy. In this way we create the same portion of CDG for each phase. The shift between the time scale in Fig. 5 and the vertical lines representing the green phases is caused by the yellow phase of three seconds. Note that at time 15[s] the platoon leader reaches maximum speed of 50 Km/h. In case of the *CDG* policy that means the whole platoon is already at maximum speed and *CDG* can fully exhibit its performance benefit. Accordingly, its throughput graph turns from a curve into a straight line. In case of *CTG*, in contrast, vehicles start moving one by one, while the *CDG* platoon is moving as a whole from the point in time when the platoon leader starts up. This is the key effect which makes *CDG* effective at traffic lights. For a more performance oriented view on the results, Fig. 6 compares the throughput improvement of all policies with the baseline *CTG-Ref* over time. While the throughput improvement of *SWITCH1ch* and *MIX* reach their saturation around 50 % around 20 s, *CDG* approaches an improvement of about 140 %. In order to illustrate the impact of different standstill distances, we ran one simulation with the double standstill distance of 6 m *(CDG-6m)*. It improved the throughput by 70%.

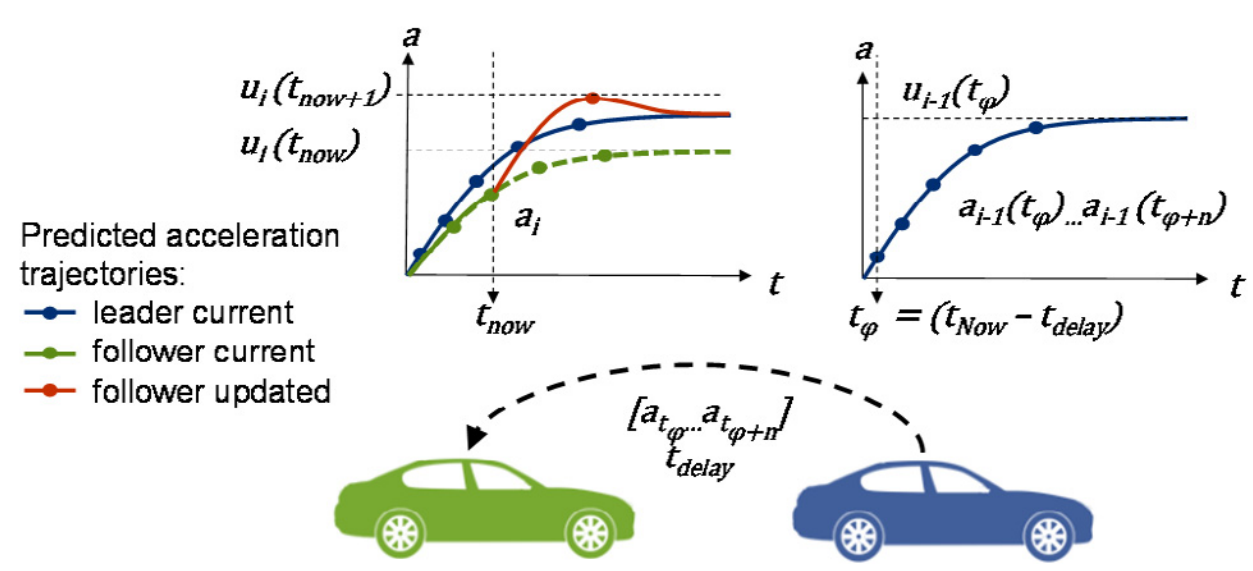

Fig. 4. Concept of the predictive CDG feed forward controller, where $t_{now}$ is the current point in time, $t_{now+1}$ is the next future point in time, $t_{delay}$ is the communication delay, $n$ is the length of the acceleration trajectory $[a_{t_\varphi}..a_{t_{\varphi+n}}]$

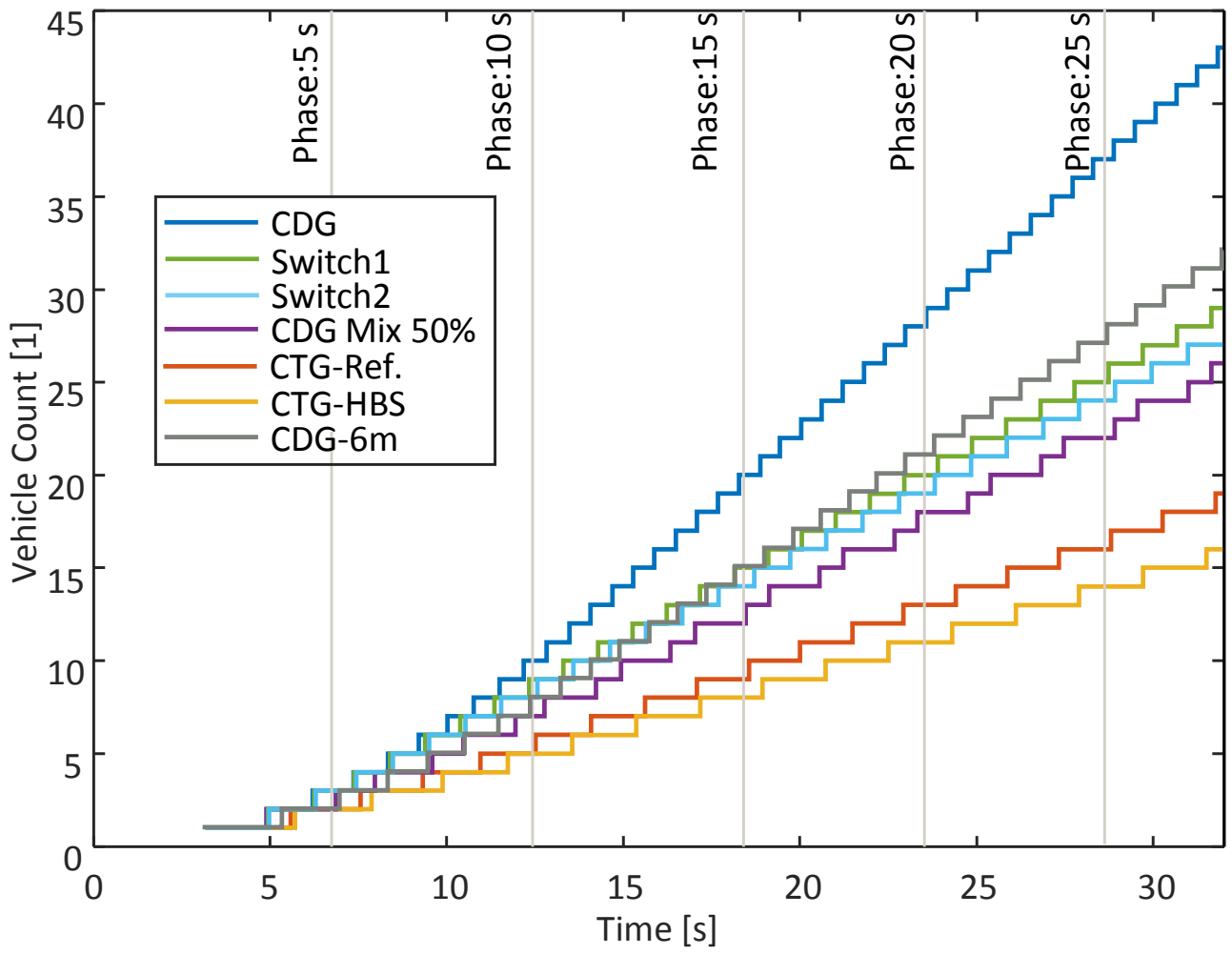

Fig. 5. Throughput comparison for the different spacing policies

### F. Conclusion

Our studies of *CDG* on start-up at a single traffic light show a performance benefit over *CTG* and the other policies. This performance benefit grows with the green phase length, and reaches 120 % at 10 s green time and then reaches a saturation of around 140 % for longer green phases. A penetration rate of 50 % *CDG* in a mix with the baseline policy only reaches 45 %, i.e. the *CDG* benefit does not scale linear with the penetration rate. In order to provide a comparison between the policies, the base parameters of the policies were calibrated on real world data and human reaction time was neglected. It must be noted, that for green phases of more than 30 s, the *CDG* platoon exceeds a length of 43 vehicles, which already could give rise to string stability issues. For that reason, for a real world implementation, counter measurements such as splitting up into mini-platoons must be considered [3], which might affect the performance. The *SWITCH1* policy, which switches from *CDG* to *CTG* at 30 Km/h reaches a performance gain of 60 %.

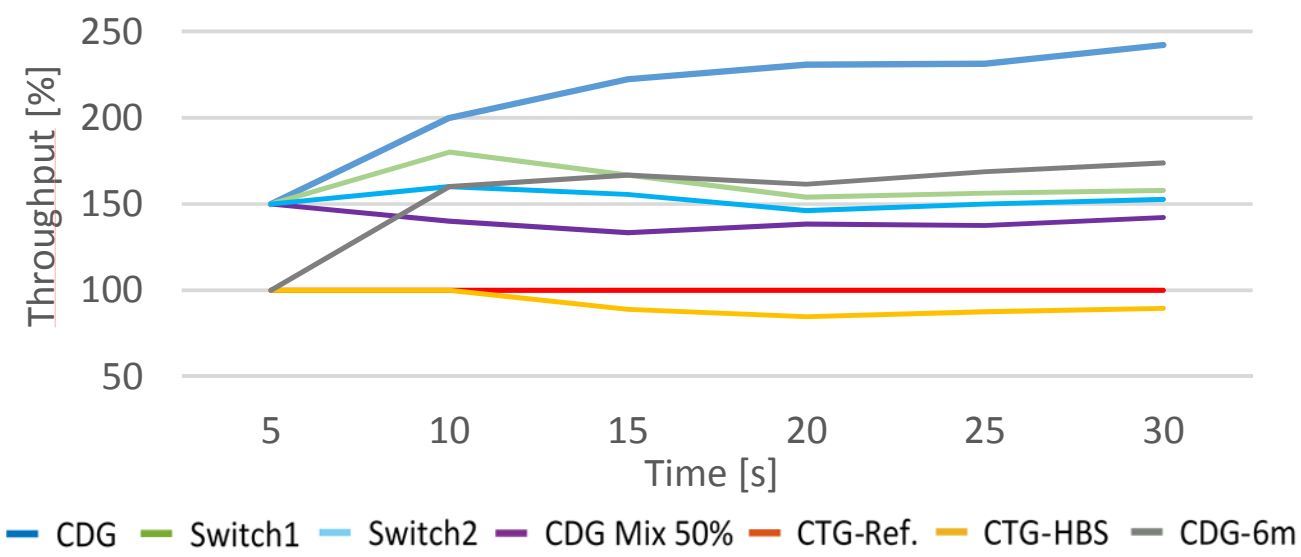

Fig. 6. Throughput improvement comparison for the different spacing policies

## IV. SINGLE INTERSECTION PERFORMANCE

In this section we expand the analysis of *CDG* from a single traffic light to a whole intersection. The performance of *CDG* policies, described in the previous section, are to a large extent due to the fact that the platoon could pass the traffic light in a free flow. However, at whole intersections the impact of traffic flow reducing factors need to be taken into account for performance comparison. This includes reduced velocities while turning, stops due to giving way while turning, as well as the fact that green light phases cannot be arbitrary long as they share the full cycle time with cross traffic and turn phases. As before, we start with the definition of an intersection layout that covers all aspects relevant for this research. Subsequently, we define further metrics to assess *CDG* performance at intersections and we finally evaluate results gained from simulating a whole intersection.

### A. Intersection Layout and Simulation Setup

Intersection layouts in urban areas include many possible constellations of elements which may have each different impact on the performance of *CDG* [27], [33]. As we have to handle and permute many parameters apart from the layout, our objective now is to define a reference layout that covers as many layout related aspects as possible and can be a fixed parameter for further studies. Note that a literature review ended up with no results on the question of what are realistic portions of left and right turning vehicles. We, thus, decide to permute both as parameters of the simulation. Fig. 7 depicts our reference layout with two lanes in each direction. Each right lane mixes straight driving with protected right turning vehicles, as there are no pedestrians. Each left lane mixes straight driving with unprotected left turning vehicles, which always need to wait for oncoming vehicles. This is ensured as there are always more vehicles waiting in front of red traffic lights from each direction than can pass it during the green phase. This oversaturation at the intersection inlets is also necessary to allow the different policies to exploit its full potential of passing vehicle per green light phase. The radius of the intersection is 20 m and turning velocity is 7 m/s which results from a maximum lateral acceleration of 2.5 m/s² [34]. Left turning vehicles entering the intersection consequently block their lane until the end of the green light phase. This reduces the random effects in the resulting throughput, independent from the desired parameterization of the simulation. We choose this particular intersection layout due to the following considerations. We should cover protected turning (turning signal phase - no yielding required) due to the reduced velocity while turning and unprotected turning (yielding required) due to its blocking effect on the following vehicles. We do not need to consider dedicated turning lanes, as they would just shift the blocking effect to occur at a higher portion of turning vehicles. We also do not need to consider dedicated traffic light phases for turning, as we already cover protected turning. We also decide to avoid lane changes in the whole scenario, in order to exclude the impact of lane changes on the simulation results. Lane changing is difficult to model, it depends on many parameters of random character, and we have no ground truth for calibration. Lane changing would further enlarge the parameter space for our simulation, while having a considerable random influence on the results. For the metrics discussed in the next subsection, missing lane changes are only relevant for the travel time of single vehicles on the blocked left lane when the right lane is free. However, we assume these to be averaged out by faster vehicles on the right lane.

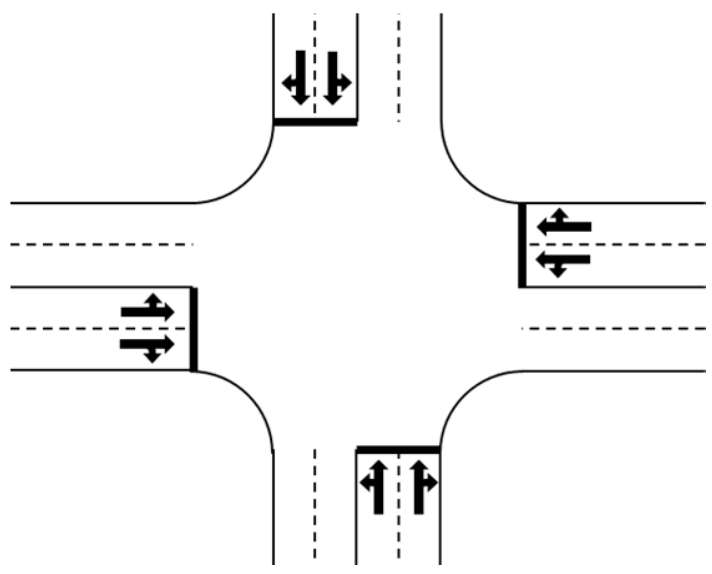

Fig. 7. Four way, two lane reference intersection layout for simulation

### B. *Metrics*

For comparing the performance of *CDG* and *CTG* at intersections, we basically measure the maximum intersection capacity [28] for both. While oversaturating the intersection inlets, we choose to measure the following metrics:

- throughput – vehicles passing per time;
- travel time – average time vehicles need to pass;
- density – portion of road meters occupied by vehicles.

Note that the German HBS and the American HCM [27], [28] define the metrics for signalized intersections based on waiting time and waiting queue length in front of traffic lights. They also consider the adjustment factors which are to be taken into account while designing signalized intersections. These metrics are not suitable for our study, as we are not aiming on optimization of configuration for the traffic light. Furthermore, measurement of waiting times contradicts our approach of oversaturating the inflows of the intersection, which is required to reveal the full potential of CDG.

Thus, we decide to apply the metrics defined by HBS for open roadways (throughput, average speed, density) and permute the configuration parameters of the traffic light setup. Instead of average speed, we measure the travel time, as it is independent from the actual travelled distance, which is difficult to be determined for random routes in SUMO [35]. The configuration parameters to be permuted are the green phase length and the ratio of left and right turns per lane.

### C. *Evaluation*

As earlier stated, our goal is not to find an optimization for the traffic light setup but to study the performance of *CDG* vs. *CTG* under all potentially occurring traffic conditions. In order to map this span of conditions, the simulation ran with 504 permutations of the following conditions, as motivated in the previous subsections:

- Intersection layout is fixed.
- Traffic flow at the intersection inlets is oversaturated, so that there are always more vehicles waiting at a red light than can pass during one green phase.
- Portion of right (0%, 10%, 30%) and left turns (0 %, 5 %, 15 %, 30 %) are permuted.
- Penetration rate of CDG and CTG are permuted with (0 %, 10 %, 25 %, 37 %, 50 %, 75 %, 100 %).
- Green light phase is permuted from 5 s to 30 s. In one permutation, the green time is the same for all directions.
- Simulation time is five full traffic light cycles.

The intersection layout and the simulation setup as described in Section IV.A, as well as the policies as described in Section III.C were implemented in the PHABMACS vehicle simulator [13]. Fig. 8 depicts a view on the intersection during simulation. The colored circles represent the radius for travel time measurement (40 m) and density (20 m). Throughput is calculated from vehicles leaving the 20 m radius per time. For the randomness to generate and equally distribute turnings and penetration ratios, PHABMACS employs the Mersenne Twister algorithm [36]. On average around 160 vehicles were in the simulation at the same time, 40 vehicles per direction.

The results of the simulations for 15 s green phase, captured in accordance with the previous section, are depicted in Fig. 9. The throughput of different *CDG* penetration rates is depicted in vehicles per hour on the vertical axis, for each permutation of left and right turn ratio on the horizontal axis. The highest throughputs were measured with no turning vehicles, at 11,550 *CDG* and 5,254 *CTG-Ref*, an improvement of 120%. The lowest throughput at 30% right turns and 30% left turns is at 4,281 *CDG* and 3,016 *CTG-Ref*, an improvement of 42%.

The improvement without turning is similar to the improvement measured at a single traffic light in the previous section. With a ratio of 30% right turns on the right lane, the improvement falls to 88% due to the reduced velocity while turning. Additional 30% left turns on the left lane almost stops the throughput on the left lane, i.e. all vehicle passing the intersection are affected by the reduced speed of the right turns, which results in a drop of the improvement to 42%.

Fig. 8. Simulation of a single intersection in PHABMACS simulator

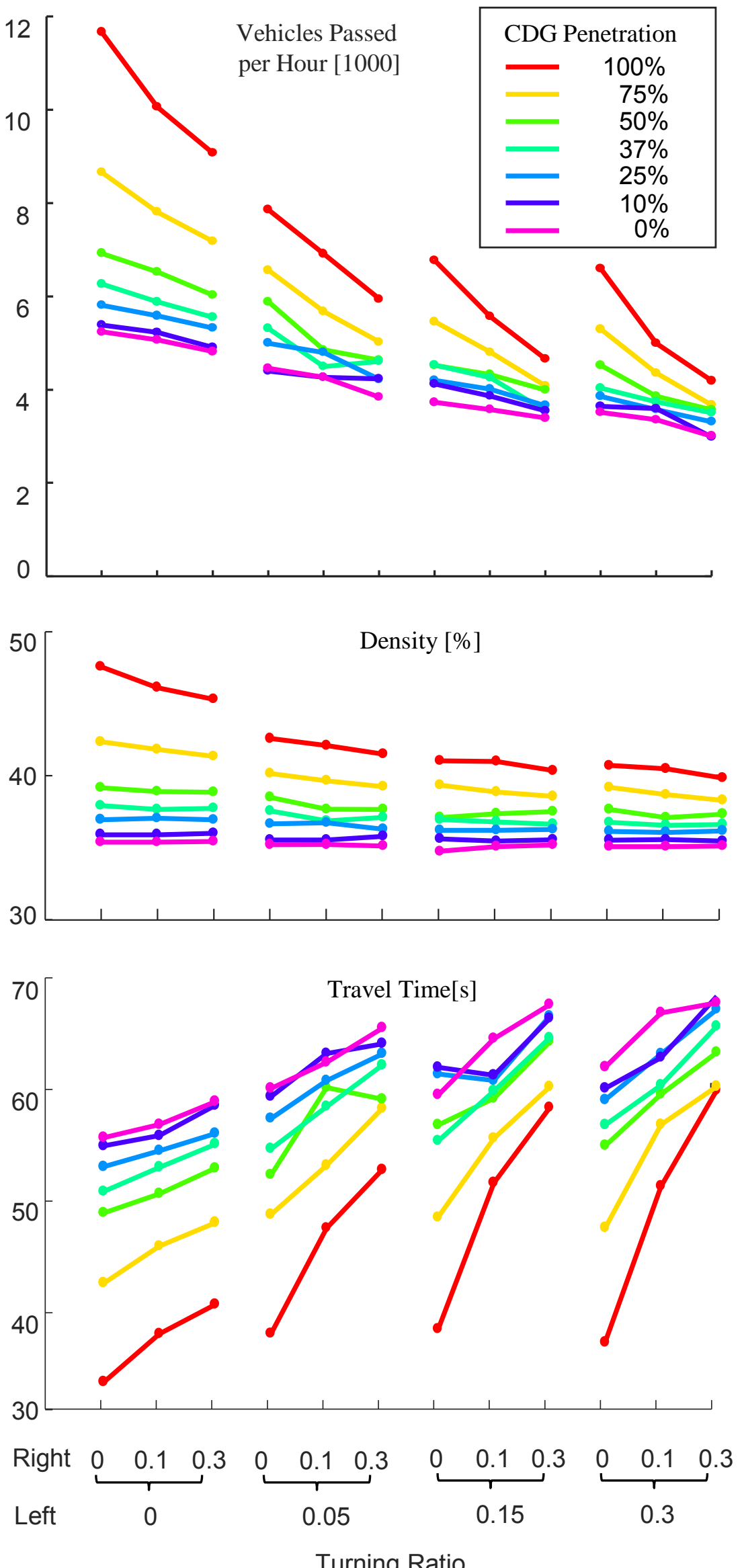

Fig. 9. Throughput, travel time, density on a single intersection CDG vs. CTH

The travel time drops from 55 s (*CTG-Ref*) to 34 s (*CDG*), which corresponds to a travel time reduction of 38 %. The lowest time reduction of 12 % results with 30 % left turns and 30 % right turns. Throughput and travel time reduction correlate with an increased density on the intersection. While the average density of *CTG-Ref* is around 35 % for all permutations, the density of *CDG* depends visibly on the turning ratios. With no turns, the density for *CDG* peaks at 47 %. For different *CDG* penetration rates, the same superlinear impact becomes apparent on throughput, travel time and density. At the first glance all graphs seem to follow an approximately uniform course. However, there are some irregularities recognizable in the pattern due to the randomness in the simulation, which affects the measured result at certain constellations during the simulation. For instance, at 5 % left turns, 10 % right turns and 50 % penetration, the throughput is the same as with 25 % penetration due to that circumstance. Moreover, although the travel time falls with an increasing *CDG* penetration, the travel time gain is of less magnitude as the throughput improvement. This is due to the fact that during the red for all directions portion of the traffic light cycle, no time benefit can be achieved by *CDG*. Discretization effects of the traffic light queue, become apparent at 15 % left turns and 25 % penetration with a higher throughput than for *CTG-Ref,* yet with a higher travel time.

Table 1 finally summarizes the throughput improvement over *CTG-ref* at all green light phases simulated. All values are rounded to the depicted number of digits. For the sake of simplicity, the table only lists the extreme values of 100 % (*CDG*) and 50 % (*Mix*) penetration with no turns, 30 % right turns only, 30 % left turns only, both at the same time. The penetration dependent improvement ratio (PIR) on the throughput is calculated by $\frac{Mix}{CDH}$ to compare the improvement of *CDG* and *Mix*. While the absolute improvement of *CDG* falls with falling green phase and increasing turning ratio, the PIR grows for short green phases and high turn ratio. At 15 s green phase length with left and right turns, the PIR peaks at 0.55 for 50 % penetration.

### D. Conclusion

In this section we broadened the study of *CDG* from a single traffic light to a whole intersection, including different ratios of protected and unprotected turning. As expected, the presence of turnings at the intersection reduced the benefit of *CDG* compared with a single traffic light. The lowest benefit was measured at 10 s green phase length, where the throughput improvement shrank from 81 % without turning to 27 % with turnings. The specific impact of turnings depends on presence and length of turning lanes. In our studies we omit such lanes in order to reduce parameter space. Thus, in our studies, one turning vehicle already blocks a complete lane.

TABLE 1
THROUGHPUT IMPROVEMENT OF CDG AT SINGLE INTERSECTION

| Turn ratios | Metric | Green Phase [s] | | | | | |
|---|---|---|---|---|---|---|---|
| | | 5 | 10 | 15 | 25 | 25 | 30 |
| no turns | CDG [%] | 50 | 81 | 120 | 134 | 140 | 135 |
| | Mix [%] | 21 | 25 | 33 | 41 | 42 | 37 |
| | PIR | 0.42 | 0.31 | 0.27 | 0.30 | 0.30 | 0.27 |
| right turns 30 % | CDG [%] | 50 | 64 | 85 | 91 | 94 | 107 |
| | Mix [%] | 21 | 19 | 25 | 28 | 29 | 28 |
| | PIR | 0.42 | 0.31 | 0.30 | 0.31 | 0.31 | 0.26 |
| left turns 30 % | CDG [%] | 45 | 64 | 94 | 92 | 112 | 109 |
| | Mix [%] | 19 | 26 | 17 | 21 | 29 | 30 |
| | PIR | 0.43 | 0.42 | 0.18 | 0.23 | 0.26 | 0.28 |
| left + right turns 30 % | CDG [%] | 59 | 27 | 42 | 50 | 45 | 46 |
| | Mix [%] | 28 | 10 | 23 | 21 | 16 | 19 |
| | PIR | 0.47 | 0.39 | 0.55 | 0.42 | 0.35 | 0.41 |

**Summary of Section IV**: *Vehicle simulations, including 160 vehicles, showed that the presence of turnings at the intersection can lower the CDG throughput improvement to 27% in worst case, compared with 140% at a single intersection. The CDG penetration rate among CTG has a nonlinear effect on its benefit. This fact is a potential hurdle for market-introduction. However, with falling absolute benefit of CDG, due to high turning rates and short green phases, the relative benefit of CDG penetration rate increases.*

## V. MODEL CALIBRATION FOR MACROSCOPIC SIMULATION

The next step for our studies on CDG is to evaluate its impact on whole traffic systems, i.e. on multiple mutually influencing intersections. As motivated earlier, development and evaluation of longitudinal control like CACC in simulation requires realistic mapping of physics. Fine differences in mapping physics and the control system interacting with its environment may lead to considerable differences to the resulting behavior. Thus, for studying CDG at a single traffic light, the sub-microscopic vehicle simulator PHABMACS is the appropriate tool (for explanations of the terms microscopic, macroscopic, and sub-microscopic simulation models see [13] or [39]). Thanks to its ability to scale out physics and control algorithms, simulating a whole intersection including hundreds of vehicles for hundreds of simulation runs is enabled [13].

However, in order to research whole traffic systems including many thousands of vehicles, PHABMACS becomes out of scope for two reasons. First, mapping that many vehicles would still require considerable time and computation capacity. Second, traffic systems under research from such macroscopic perspective may also produce realistic results, provided that an appropriate model is leveraged, which maps the microscopic behavior sufficiently in a macroscopic scale.

In the following, we propose a methodology to calibrate and validate a sub-microscopic simulation model against a microscopic simulation model, in order to enable macroscopic traffic analysis including several thousand vehicles. We use this methodology to match the implementation of CACC controllers in PHABMACS and its validated vehicle model to the SUMO [38] traffic simulator. Calibration and validation are essential here in order to ensure that the traffic simulation model in SUMO generates the same results regarding relevant metrics as vehicle dynamics simulation model in PHABMACS.

### *A. Model*

In order to map CACC in SUMO, we choose the Krauß car-following model [39] as the basis implementation. The model is directly applicable for CTG. For CDG, however, we need an adaption of the model, as fixed following distances cannot be realized for the following reason. Although, the Krauß model has a parameter for the velocity dependent time gap, setting this parameter to 0 s does not make the vehicles start up at the same time. Each vehicle starts exactly one simulation time step later than its predecessor. Since all vehicles follow the same acceleration trajectory, the inter vehicle distance is constantly growing while accelerating and shrinking while decelerating.

For this reason we modified the Krauß model according to (1). As with the Krauß model, we base our model on the calculation of a maximum safe speed $v_{safe}$. If the distance $s$ to the predecessor is greater than the standstill distance $s_0$, we apply the Krauß model with a small modification. The tolerance band $s_t$ is added to $s_0$ for the calculation. In this way, a tolerance band around $s_0$ is created. This allows the vehicle to overshoot the stand still distance by $s_t$, which is required as a buffer for driving with constant distances. If the distance $s$ is within this tolerance band, $v_{safe}$ is set to the predecessors velocity $v_l$. If the band is undershot, $v_{safe}$is set to the $v_l$ reduced by a factor $d$ (0.95), to make the vehicle return to the tolerance band. The simulations step size needs to be aligned with $s_t$, in our case $s_t$ equals 0.5 m at a simulation step size of 0.1 s.

$$v_{safe} = \begin{cases} v_l & s_0 - s_t < s < s_0 \\ v_l d & s < s_0 - s_t \\ -b\tau + \sqrt{b^2\tau^2 + v_l^2\ 2b(s - (s_0 + s_t))} & s > s_0 \end{cases} \quad (1)$$

This looks like a hack of the car-following-system designed for SUMO and we would recommend to use this model for specific applications only. However, for our use-case, it works sufficiently good, as demonstrated in the next subsection. Another required modification, is to enable followers to catch up with their predecessors who drive with maximum speed. For this purpose, we lowered the maximum speed of vehicles without predecessors to 95% of the speed restriction of the current link in SUMO. This is also done for the Krauß model.

### *B. Calibration and Validation Method*

Our proposed validation methodology consist of two steps. First the models of both simulations, vehicle simulation (PHABMACS) and traffic simulation (SUMO) are calibrated. This calibration aims to the match of time and location of each vehicle during the simulation for the same scenario in both simulators. Second, the metrics determined to evaluate simulation results, are determined in both simulators for the same scenario and validated against each other. This model validation method was designed following the consideration of balance between effort and value of model confidence, presented in [13]. Accordingly, this method does not aim at finding the limits of model validity, but to assure validity of the considered simulation scenario to generate valid metrics.

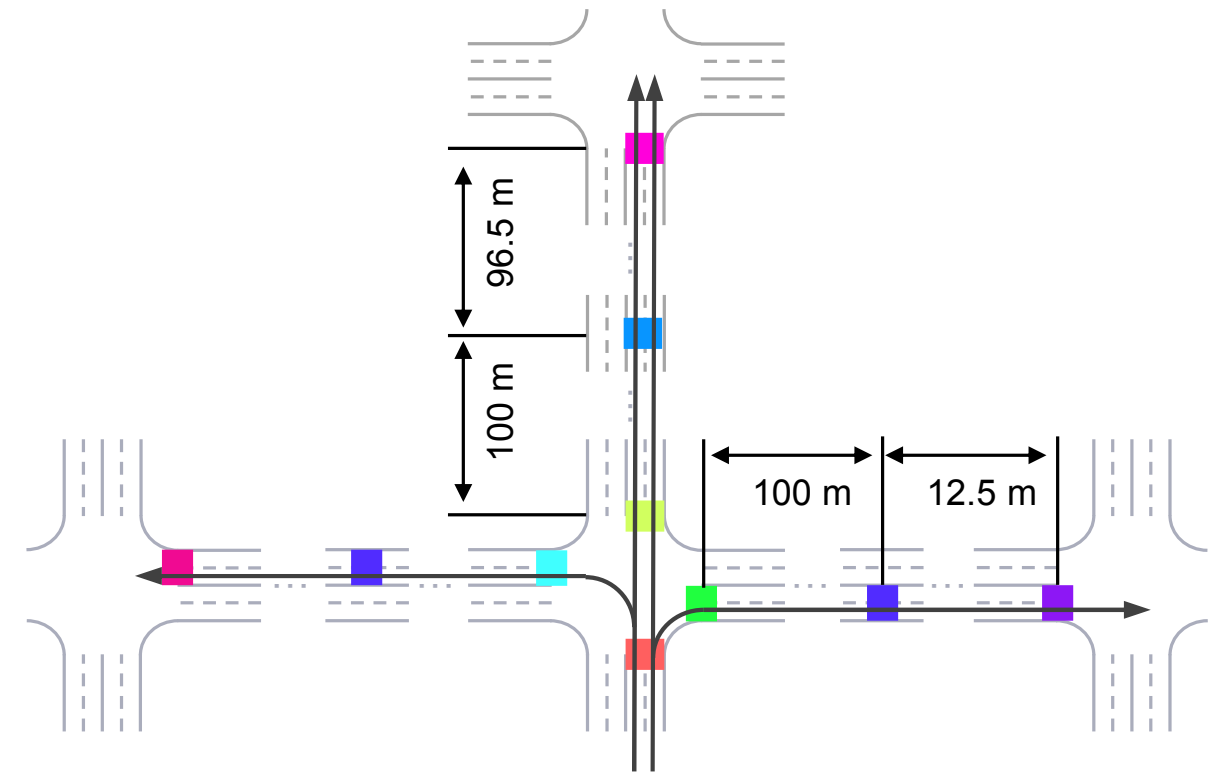

Fig. 10. Calibration setup SUMO - PHABMACS

*1) Step I Timing*

In order to calibrate the timing and location of a vehicle (time-space domain), we first need to place detectors in both simulations at crucial, scenario specific locations. In our case our objective is to consider multiple mutually influencing intersections, i.e. our scenario includes one central intersection and one adjacent intersection in each direction as depicted in Fig. 10. As the intersection layout is identical from each direction, we just need to regard vehicles incoming from one direction at the central intersection. The distances between the intersection, as well as the traffic light cycle times and their offset between the intersections are chosen in accordance with the next section. The detectors are placed according to Fig. 10. In this way, start-up characteristics, travel time through and between intersections are covered. Again, lane changing is disregarded for the aforementioned reasons. Left turns stop the traffic on the left lane and the turning vehicle is the only one passing the traffic light for the current cycle.

In order to validate that timing in both simulations is similar, we run both simulations for all relevant permutations of simulation parameters and compare the vehicle counter for all detectors. This is performed automatically, so that the shortest cycle time possible of manually tuning the model parameters and validation is achieved. We need to assure that for all permutations of CDG penetration, traffic light cycle times and offsets, the correct number of vehicles pass per traffic light cycle. As the shortest traffic light cycle time to be studied is 5 seconds, a limit for the maximum difference between corresponding detectors in both simulations of 1 second is sufficient.

For the assessment of validity, we propose the objective timing criterion as described above, complemented by a subjective criterion as motivated in [13]. If a simulation scenario ran invalid and the number of detector that show higher differences than 1 second is small, the verdict of validity can be changed manually if reasonable. One example for such a subjective verdict is depicted in Fig. 11. While the objective criterion can be applied automatically, the subjective criterions needs to be assessed manually. The idea here is to apply automatization to the greatest extent, while reducing manually effort to assess the edge cases. For further information on this methodology, please refer to [13].

The count on each detectors is depicted with the corresponding color of Fig. 11. The simulation ran at a traffic light cycle of 15 s, no offset between intersections, with vehicle queue from south of: 6 left (*CTG*), 6 right (*CTG*), 18 straight (*CTG*), 8 left (*CDG*), 8 right (*CDG*), 21 straight (*CDG*). Around simulation time 150 s, a slight higher compactness of the *CTG* platoon in SUMO causes a time difference at the intermediate straight detector of 1.2 s. Around time 239 s, the *CDG* platoon of 20 vehicles stops in front of the north intersection. The 15th vehicle stops right on the same detectors in PHABMACS, while in SUMO the corresponding vehicle stops slightly in front of the detector. Thus, a time difference of a full cycle time is measured. The final parameterization (notation according to [42]) of the SUMO model after calibration is listed below.

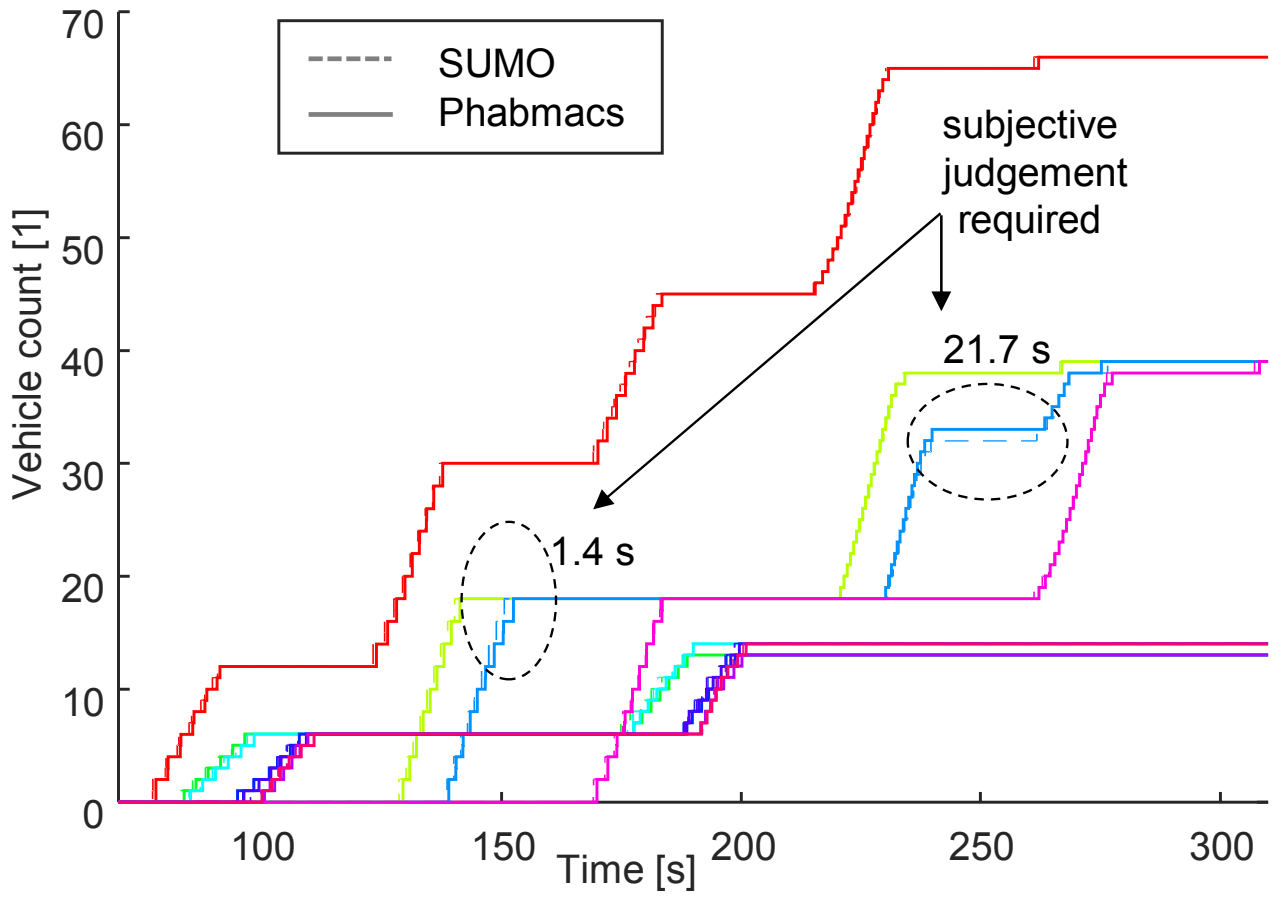

Fig. 11. Subjective validation criterion – example: green light 15 s, offset 0 s

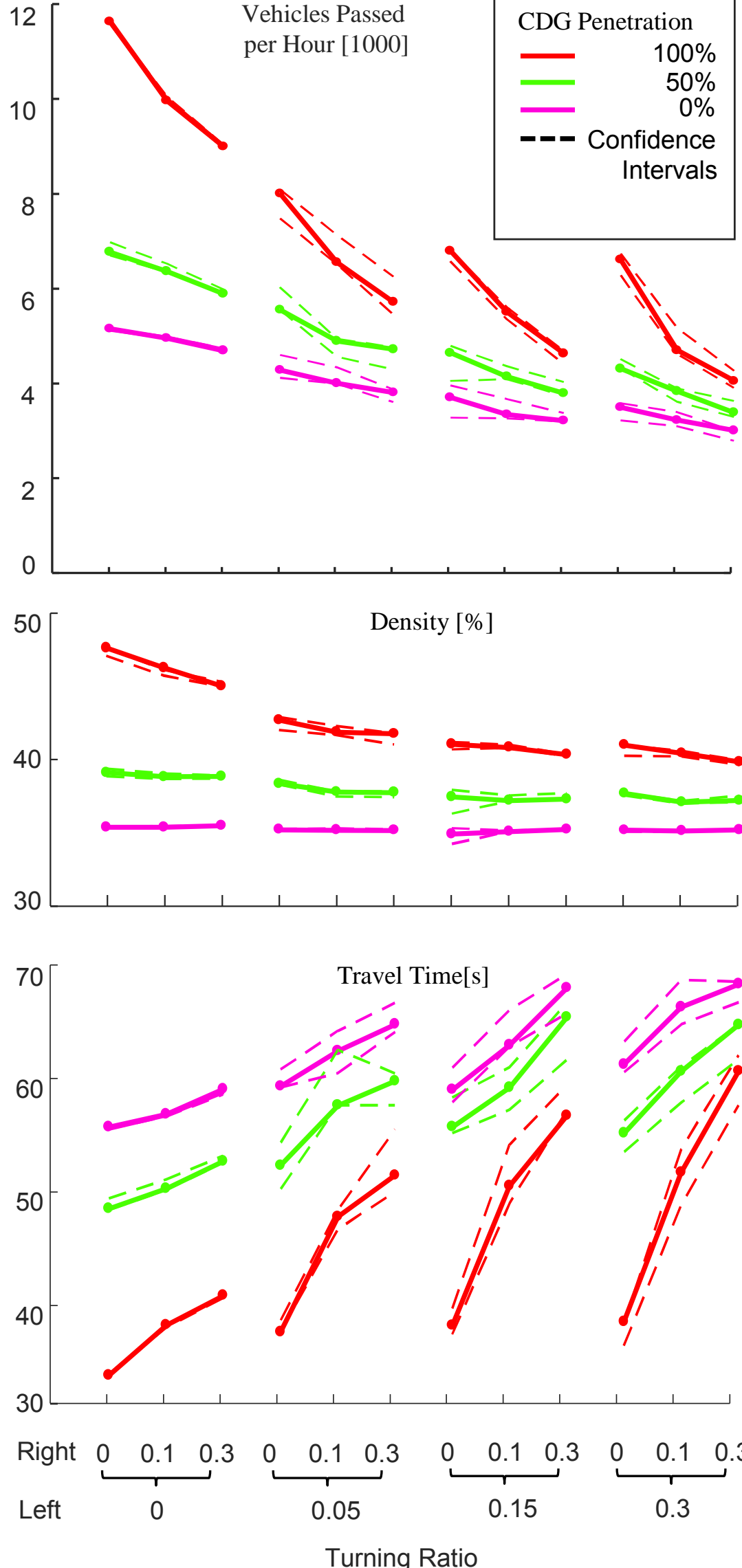

Fig. 12. CDG model metrics validation SUMO - PHABMACS

TABLE 2 SUMO MODEL PARAMETERS

| Model | Parameter in SUMO | | | | |
|---|---|---|---|---|---|
| | decel ($\boldsymbol{b}$) | accel ($\boldsymbol{a}$) | tau ($\boldsymbol{\tau}$) | minGap ($\boldsymbol{s_0}$) | Sigma |
| CTH | 4.70 | 1.70 | 0.9 | 2.95 | 0.4 |
| CDG | 4.70 | 1.40 | 0.02 | 2.45 (+ $s_t$ 0.5) | 0.02 |

### 2) *Step II Metrics Validation*

In step II, the calibration completed in step I is validated. The approach of our proposed validation method is based on the statistical analysis of the same simulation scenario in both simulators. By proceeding in a similar fashion to the validation of a sub-microscopic simulation model against a real world vehicle in [13], we validate a microscopic traffic simulator (SUMO) against a sub-microscopic vehicle simulator (PHABMACS). As described in [40] we employ the 95 % confidence interval of the relevant metric measured at multiple simulation repetitions for analysis. The metrics to be validated in this case are throughput, density, and travel time. The confidence intervals for each metric depicted in Fig. 12. CDG model metrics validation SUMO - PHABMACS for the 15 s phase time simulation run corresponding with Fig. 9. The confidence intervals are determined as described in [13], using the MATLAB® implementations of the Student's t inverse cumulative distribution function "*tinv*", and the standard deviation "*std*" for $\sigma$, where $v$ is the degree of freedom (the number of simulations, six in this case) and $\mu$ is the mean value of data.

$$U, L = \left\{ \mu \mp C \frac{\sigma}{\sqrt{N}} \right\},\ C = tinv(0.95, v),\ N = \frac{v}{2} - 1 \qquad (2)$$

All 504 permutations (see previous section) were simulated six times in PHABMACS and in SUMO. We consider validity as achieved if the average metric measured in SUMO is inside the confidence band measured in PHABMACS, which is the case as shown in Fig. 12.

## VI. MULTI INTERSECTION PERFORMANCE

In this section, we analyze the crucial traffic hindrance situations caused by CDG, which lead to a decreased performance of CDG compared with the single intersection analyzed earlier. Two main factors lead to such a lowered performance. First, congested intersection outlets that lead to obstructed off-flowing traffic, and second, reduced in-flowing traffic. Both can be caused by the influences of the overall traffic system. Thus, the main question to be discussed in this context is the impact of CDG on the traffic system, or more precisely on multiple mutually influencing intersections. To proceed, for consistency with our previous discussion, we apply the same intersection layout. We combine this layout to two synthetic simulation scenarios, an arterial signalized corridor [41] with five intersections and a coordinated grid network [41] of 25 intersections. Including up to 5500 vehicles per run, both scenarios are simulated with multiple permutations of traffic light configurations and turning ratios. With regard to simulation runtime we can afford such a number of vehicles and this span of permutations, thanks to the calibration of the CDG and CTG model with the traffic simulator SUMO. The results are calculated counting vehicles entering/leaving the simulation 20 m far from the outer intersections, i.e. excluding the unbound queues. All trips end outside this area. Final values are captured when all metrics increased to a steady state level and keep it for five hours simulation time. For all simulation scenarios, we assume a maximum possible traffic inflow and an unobstructed outflow of the traffic system. Finally, we model a real world road network simulation scenario using a real world traffic layout and traffic light configurations in the subsequent section.

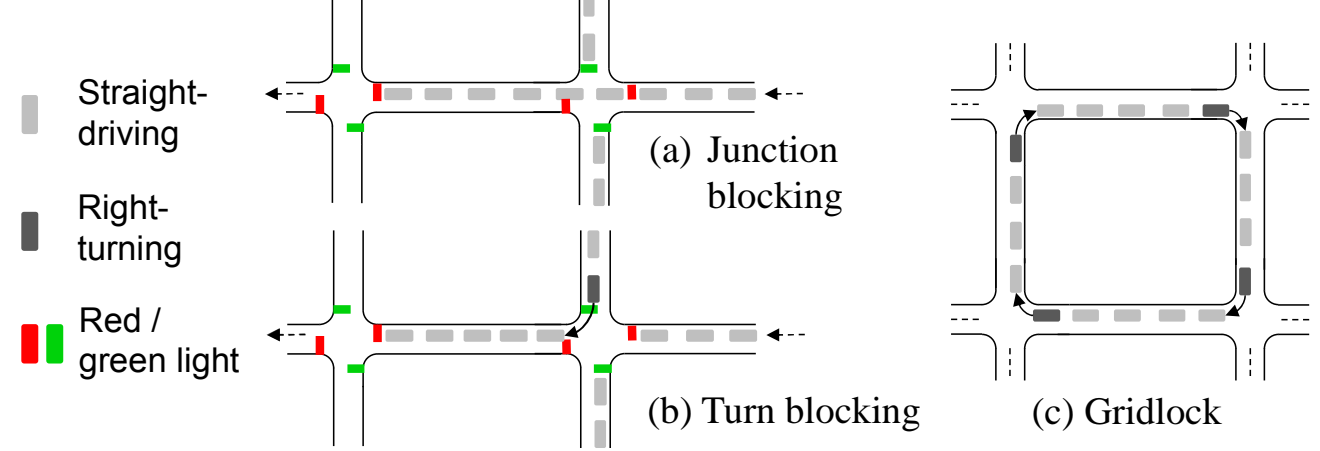

Fig. 14. Traffic hindrance situations

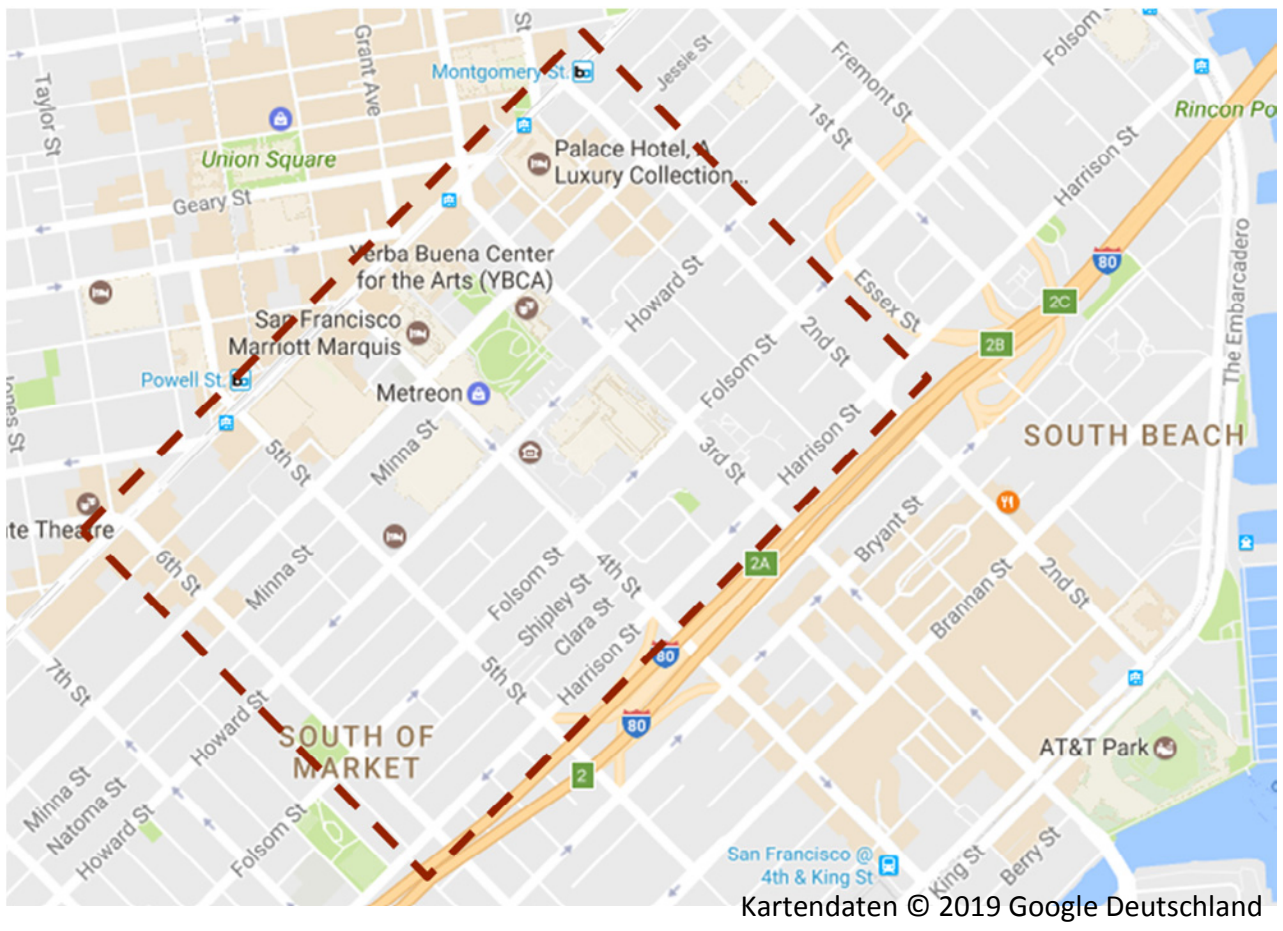

Fig. 13. Simulation scenarios arterial and grid based on urban are in, San Francisco (South of Market), CA, between Market St. and Harrison St.

The topology of the arterial scenario and the grid scenario is based on an urban area in the South of Market neighborhood in San Francisco (see Fig. 13). We choose this area as it has an even grid of intersection all with the same distance of 276.5m NW bound and 192.5m SE bound. These constant intersection interspaces enable isolating the impact of interspace length on the simulation results from the other simulation parameters. Although this area partially consists of one-way streets in real world, we unify the simulation scenario with two way streets and intersection layouts according to Section IV.A.

### A. *Traffic Hindrance Situations*

Applying CDG at the constellation of intersection in the described way, leads to three traffic hindrance situations. The disturbance effects resulting from these three situations depicted in Fig. 14 are described in the following.

#### 1) *Situation 1 – Junction Blocking*

Assuming the traffic backlog from a traffic light reaches the adjacent intersection as shown in Fig. 14 (a). Under certain circumstances vehicles come to a stop on the middle of the intersection and do not leave before the traffic light switches to the phase for the cross traffic. In this situation, the cross traffic has to wait for a full traffic light cycle until the intersection is clear again. Due to the close distances in a CDG platoon and the one-vehicle look-ahead pattern, this event occurs more often than with CTG. CTG by its very nature creates a contraction of the platoon while stopping and thereby more space on the intersection area. In order to create spaces on the intersection, CDG would require a coordination between vehicles, such as described in [37]. In SUMO there is a heuristic mechanism (*no-block-heuristic*) that helps vehicles to anticipate a possible hold at a position which blocks the cross traffic. However, as in the real world, in some specific situations, this predictive mechanism does not always work out.

*2) Situation 2 – Turn Blocking*

Even if vehicles stop to prevent a junction blocking, traffic backlogs might prevent vehicles from turning. In this case, as depicted in Fig. 14 (b), the cross traffic behind the turning vehicle is blocked for the current traffic light cycle. This applies for right and left turning vehicles. This event is also more likely to happen with CDG than with CTG for the aforementioned reasons.

*3) Situation 3 – Gridlock*

If situation 2 occurs at four intersections at the same time, this leads to a complete standstill beyond subsequent traffic light cycles (see Fig. 14 (c)). For such situations, SUMO offers a mechanism (*teleport* [42]) to model the real-life behavior of eventually finding a way around the blocking vehicle and so resolving the gridlock. For all experiments described in the following subsections, we set the waiting time in SUMO for each vehicle to resolve gridlocks and turn blockings to three full traffic light cycles. Solving junction blockings is set to the time of two green light phases.

## B. *Arterial Signalized Corridor – Simulation Scenario*

The arterial scenario consists of five adjacent intersections of a major street with a distance of 192.5 m, as depicted in Fig. 15. The two lane layout of sections IV.A, as depicted in Fig. 7, is applied. The arrows in the Fig. 15 mark the high traffic inflows. As described earlier, lane changes are suppressed in order to exclude the impact of a lane change model on the simulation results. This scenario represents coordinated intersections on a major street. Thus, the green light portion of the cycle time is longer for the major street than for the minor streets. The following parameters were applied for the simulation:

- turning rates on minor roads: left 20 %, right 40 %;
- turning rates on main road is permuted with two different parameterizations: 1 (no turning), 2 (left 10 %, right 20%);
- penetration rates are permuted with 0 % (*CTG*), 50 % (*Mix*), and 100 % (*CDG*);
- green light portion for the major street is permuted with 25 s, 30 s, and 35 s with corresponding 10 s, 7 s, and 5 s for the minor streets;
- offset time (time shift between the traffic light cycles) between intersections is permuted with 0 s and 15 s.

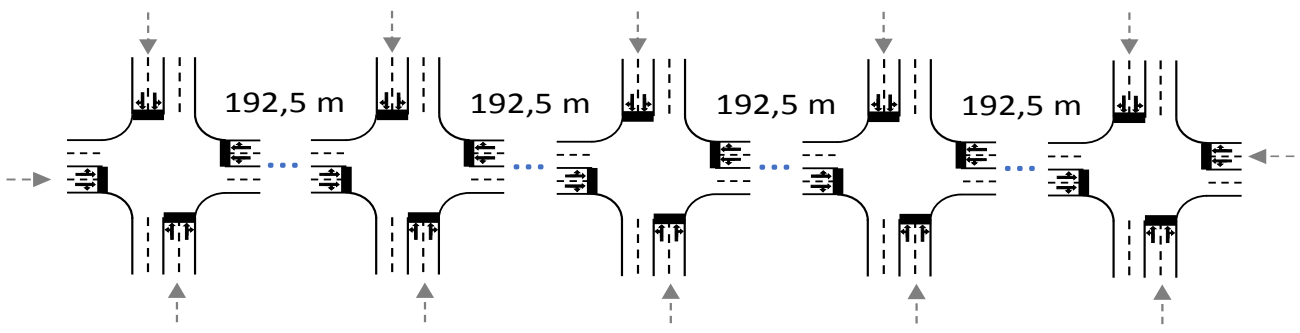

Fig. 15. Layout arterial signalized corridor with five intersections

*1) Impact of Green Phase and Offset Time Between Coordinated Intersections for CDG on Arterials*

For a better understanding of the arterial scenario simulation results, we present some preliminary considerations in the following. The performance of CDG in such a scenario is heavily influenced by the ratio of platoon length and intersection interspace. Assuming that there are no turnings and lane changes, the platoon length is indirectly controlled by the green light phase. Fig. 16 depicts four different situations to be distinguished regarding the named ratio:

*Situation 1* – Fig. 16 (a) - Platoon length (81.5 m at 10 s green phase) is shorter than intersection interspace and traffic lights are synchronized, i.e. of same cycle time and no offset between their cycles. After starting up, the platoon needs about 15 s to travel to the next intersections at 50 Km/h. However, as the full cycle time is 36 s, additional waiting time at the next intersection results in a travel time of 36 s per intersection.

*Situation 2* – Fig. 16 (b) - Same parameters like Situation 1 with an additional offset between the traffic light cycles of 15 s (from left to right in Fig. 16). This offset reduces the travel time to 15s per intersection in one direction, as the platoon does not need to stop. In the opposite direction, the platoon still needs to stop, however, the waiting time is reduced by 15 s to 21 s. This results in an average travel time for both directions of 18 s per intersection. This shows that synchronized traffic lights are always the worst case in terms of travel time. Any offset has a positive impact.

*Situation 3* – Fig. 16 (c) - Platoon length is longer than the intersection interspaces (in our case for green times longer than 15 s). The platoons stopping at a traffic light protrude into the adjacent intersection, which leads to the traffic hindrance situations junction blocking and turn blocking as described in the previous subsection. This leads to a falling traffic throughput and an increased travel time compared with Situation 1 and 2.

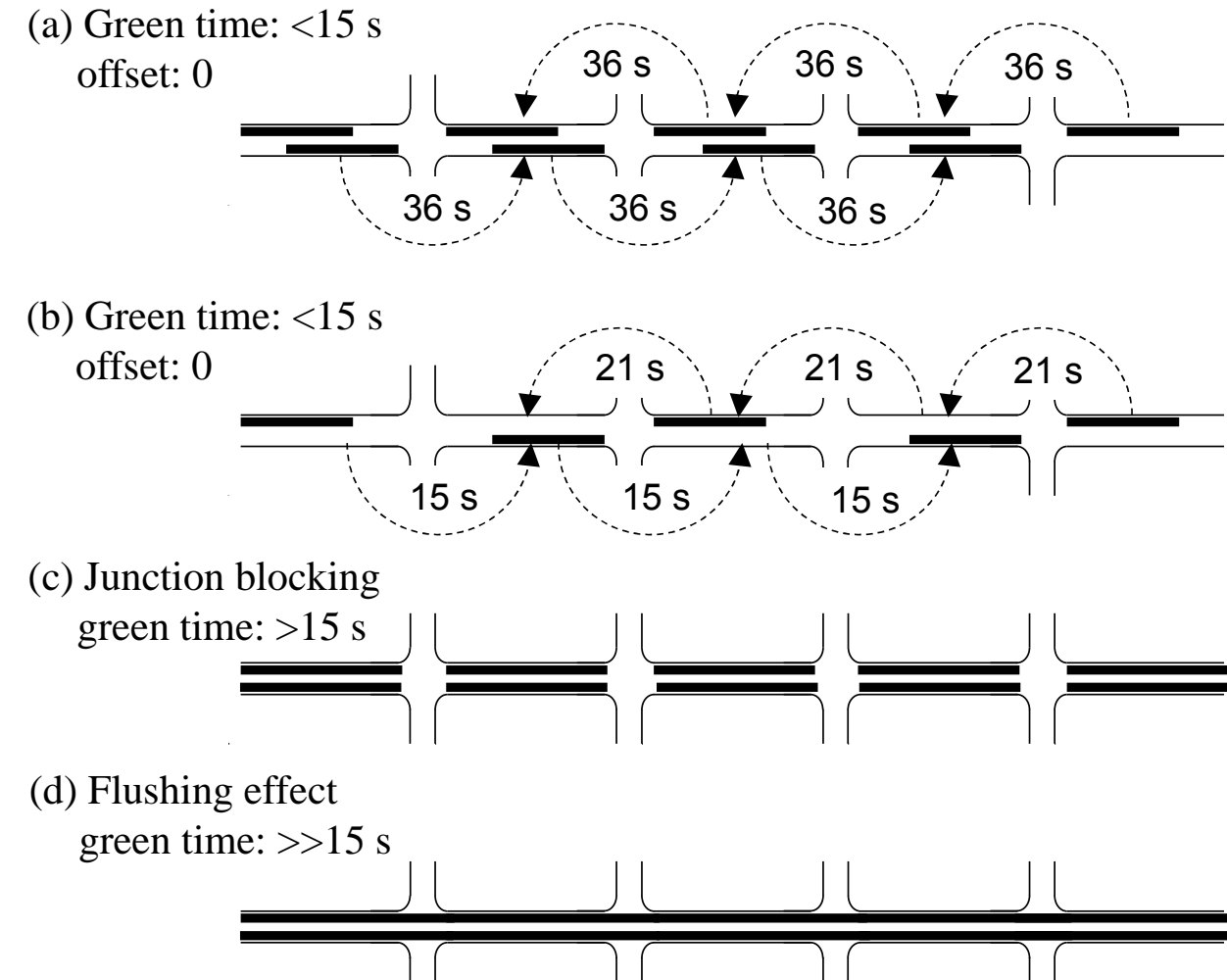

Fig. 16. Impact of green phase and offset time between coordinated intersections for CDG on arterials

*Situation 4* – Fig. 16 (d) - Relatively long green times on the major road lead to a platoon length which spans multiple intersections and results in a flushing effect. While junction blocking still occurs, its negative effect on throughput and average travel time is compensated by the flushing of traffic. The throughput increases due to the short red time (long green time) portion on the major road and the travel time falls as vehicles don't need to stop at each intersection.

*2) Results*

Fig. 17 depicts the simulation results without turnings on the major road. Fig. 18 depicts the simulation results done with 10 % left turnings and 20 % right turnings on the major road. In both figures, sub-figures a, b, c depict the throughput, travel time and density measured for *CDG*, *CTG*, and *Mix*. Sub-figures d, e, f depict the improvement of *CDG* and *Mix* over *CTG*. In each sub-figure the relevant metric is plotted at the vertical axis on a ground plane which represents the permutation of green time and offset.

*3) Discussion of results without turnings on the major street*

**Throughput -** Fig. 17 (a) and (d): *CTG* and *CDG* throughput both increase with green time length, while an offset has a slightly negative effect on both above 30 s green time. *CDG* shows an improvement of around 50 % in average, while *Mix* is around 35 %. This means that in contrast to a single traffic light scenario, the *CDG* improvement for this scenario scales better than linear with the penetration rate. This is due to the fact, that disturbance effects resulting from the named traffic hindrance situations have a higher negative influence, the higher the *CDG* penetration rate is. The overall improvement is lower than for a single intersection, since all green times simulated are above 15 s, which means in all cases the disturbance effects junction blocking and turn blocking occur.

**Offset impact on throughput -** The negative impact of the offset on the throughput above 30 s green time can be explained as follows. The longer the green time, the higher is normally the negative impact of disturbance effects, due the resulting increased the cycle times and, thus, less opportunities per time to clear the intersection. The flushing effect compensates this negative impact, as its positive impact increases with the green time length. The flushing effect, however is negatively influenced by the offset, as it reduces the time when all intersections are green coevally. All these facts together lead to throughput minimum of *CDG* at 30 s green time and 15 s offset.

On the other hand, the offset has a positive impact on the throughput of the minor streets. It mitigates the junction blocking problem, as the intersection is always cleared in one direction due to the shifted red time at the adjacent intersection. However, due to the high portion of left turnings on the minor streets in our simulation, this positive impact is of low significance and, hence, the negative impact of the offset prevails in the resulting throughput.

A separate simulation without turnings on the minor streets, not depicted in the figures, resulted with a *CDG* throughput improvement of 65 % at 25 s green time and 85 % at 35 s green time. We measured the same throughput with and without offset for each green time length. This simulation revealed that without turnings, the positive offset impact on the junction blocking could completely compensate the negative offset impact on flushing.

Fig. 17 Arterial scenario simulation results without turnings

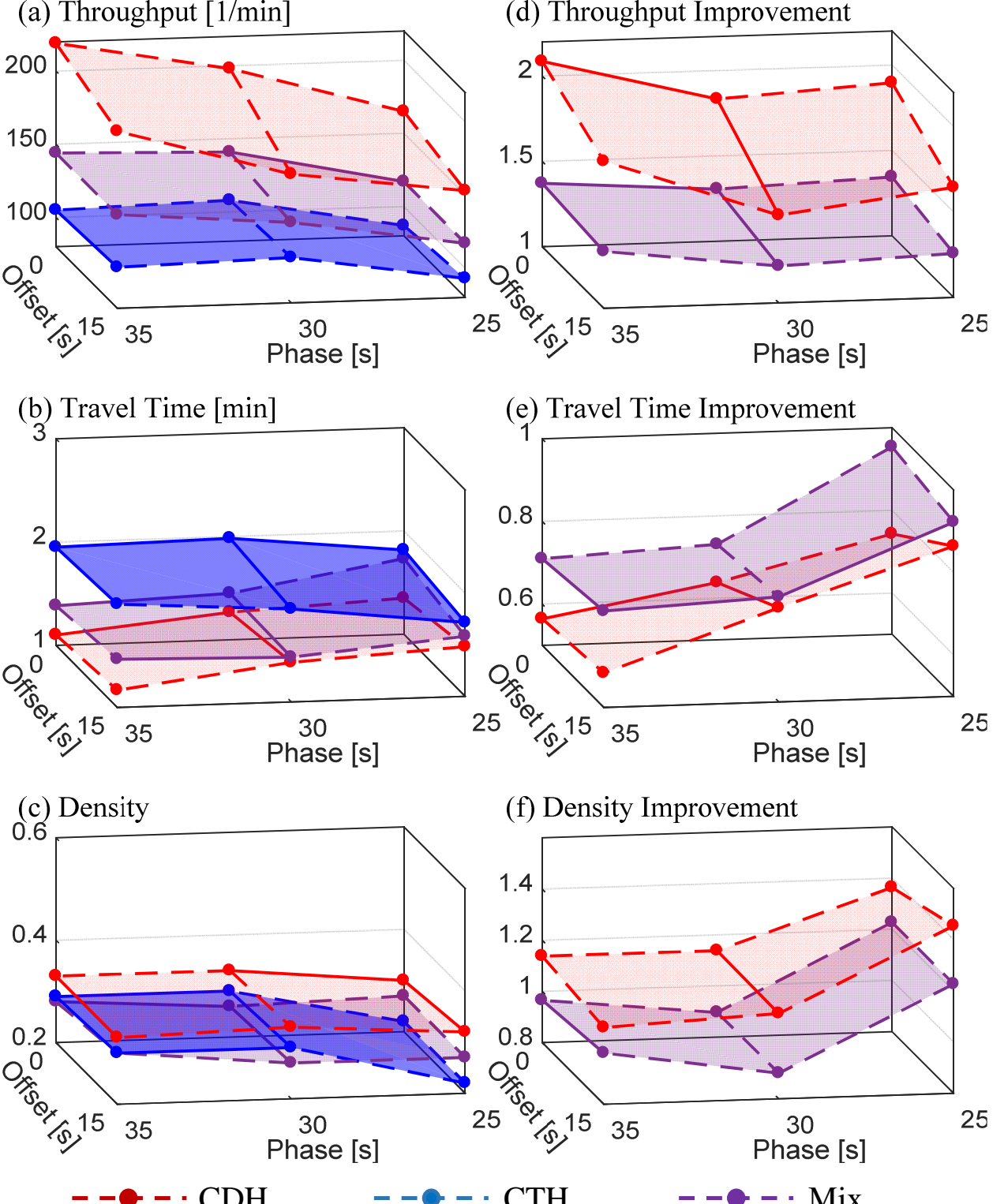

Fig. 18. Arterial scenario simulation results with turnings

**Travel time** - Fig. 17 (b) and (e): While the travel time of *CTG* is approximately equal for all green times and offsets, *CDG* travel time notably benefits from offset. The travel time improvement of *CDG* and *Mix* over *CTG* are both around 20 % without offset and around 30 % with offset. In general, high green times have a positive impact for both. The offset related difference correlates with the lower throughput and density of *CDG* with offset. Here, *CDG* increases the throughput at the expense of density and travel time. Another indicator for this relationship is shown by the fact that throughput of *CDG* is higher than *Mix*, while their travel times are almost equal. The most remarkable permutation regarding travel time is at 25 s green time without offset. This permutation results in the highest travel time for *CDG*, mainly caused by the turn blocking problem. Turnings from the minor streets cannot enter the main street, which is notably mitigated when offset is present and compensated in average by the flushing effect at higher green times.

#### 4) Discussion of results with turnings on the major street

**Throughput -** Fig. 18 (a) and (d): A notably increased throughput for *CDG* can be seen in the results with additional turnings present on the main street. While the *CTG* throughput is in saturation at 30 s green time, the *CDG* throughput increases linear with the green time length. Its improvement over *CTG* peaks at 110 % at 35 s green time. This difference is caused by the vehicles leaving gaps on the main street platoons when turning. In this way the platoons can contract at red lights, which mitigates the junction blocking and turn blocking effect. *CTG* on the other hand is negatively influenced by the turnings, especially at longer green times. The offset shows the earlier explained influence on the flushing effect for *CDG*. Its positive impact on blocked turnings at minor streets does not come into effect, as the gaps on the major street already mitigate turn blocking.

**Travel time** - Fig. 18 (b) and (e): Introducing turnings on the major street results in halving of the density in simulation for all policies. This leads to an overall reduced travel time. The travel time improvement of *CDG* ranges from 20 % to 45 %, while *Mix* goes in saturation around 30 % at 30 s green time. Here, *CDG* travel time is not affected by the offset, while there is a slightly negative impact on *CTG*.

#### 5) Arterial Signalized Corridor Simulation Results Summary

From the simulation results we observe the following facts about *CDG* applied in traffic system (specifically at adjacent and mutually influencing intersection on arterial streets).

- In contrast to single (or isolated) intersections, the high traffic density caused by the *CDG* platoons may lead to the disturbance effects, junction blocking and turn blocking.
- These effects lower the room for improvement of *CDG* over *CTG*. Lower penetration rates (*Mix*) are less vulnerable to these effects, which increases their relative benefit.
- The overall high density on the major street is usually mitigated by turnings leaving gaps in the platoons on the major street.
- Additional countermeasures to lower the density are offsets and green times which create platoons shorter than intersection interspaces.
- Long green times that entail platoons spanning multiple intersections cause a flushing effect in the major street that improves throughput and travel time, however vehicles on minor streets still suffer from disturbance effects.
- Offsets in general reduce travel time for CDG and can reduce disturbance effects in one direction, however lower the flushing effect.

### C. Grid Scenario

The grid scenario includes all 25 adjacent intersections marked in Fig. 13. Again, the two-lane layout of section IV.A as depicted in Fig. 7 is applied and lane changes are suppressed. Maximum possible traffic inflows are specified at the 20 inlets. This scenario represents a coordinated grid network [41] of intersections that connect major streets. Thus, the green light portion of the cycle time is equal for both directions. The following parameters were applied for the simulation:

- turning rates are permuted with two different parameterizations: 1 (no turnings), 2 (left 5 %, right 10 %);
- penetration rates are permuted with 0% (*CTG*), 50% (*Mix*), and 100 % (*CDG*);
- green light portion is permuted with 5 s, 10 s, 15 s, 20 s;
- offset time between intersections is permuted with 0 s, 5 s, 10 s, and 15 s.

#### 1) Results

Figure Fig. 19 depict the grid simulation results done without turnings and Fig. 20 with 5 % left turnings and 10 % right turnings. The sub-figure structure is similar to Fig. 17 / Fig. 18.

#### 2) Discussion of results without turnings in the grid

**Throughput** - Figure Fig. 19 (a) and (d): *CTG* and *CDG* throughput both increase with green time length. While *CTG* goes in saturation at 15 s green time, *CDG* shows a dent at 15 s, which is caused by the junction blocking beginning on the shorter axis of the grid. At the longer axis, junction blocking occurs from 20 s green time on, however, in sum we see a further increasing throughput. The improvement of *CDG* has its maximum of 100 % at 10 s green time and approaches 70 % above 20 s. This value matches the results of the arterial scenario without turnings on the minor streets (not depicted in the figures). The offset has no notable influence on all policies.

**Travel time** - Fig. 19 (b) and (e): The travel time increases for *CTG* from 2.5 min at 5 s green time to 3.3 min at 20 s green time, due to the increased cycle times. Up to 10 s green time *CDG* saves travel time as expected, while above 15 s green time, the junction blocking lead to a considerably increasing traffic density and, thus, to an increased travel time. Here, throughput is increased to the expense of travel time again. In contrast to the arterial scenario, junction blocking effects both directions. Thus, the average travel time is affected in both directions by many vehicle which need to wait two cycles at the same intersection. *Mix* has no significant travel time improvement and the offset has a positive impact in all policies.

*3) Discussion of results with turnings in the grid*

**Throughput** - Figure Fig. 20 (a) and (d): *CTG* shows similar characteristics to the case without turnings but with approximately 20 % lower throughput. For *CDG* the throughput drops significantly at 15 s green time. This drop results from gridlocks occurring in addition to the junction blocking and turn blocking as discussed earlier. While *Mix* shows an average throughput improvement to about 20 %, *CDG* drops from 60 % to 20 % at this significant threshold. Without offset, *Mix* even generates a higher throughput than *CDG* at 20 s green time. Offset shows an overall positive impact on *CDG* as it creates free spaces and so counteracts gridlocks. In contrast to the arterial scenario, longer green times don't lead to platoons spanning multiple intersections, i.e. intersection are not cleared which contributes to arising gridlocks. Rather, gridlocks are a local. This becomes apparent by observing the traffic density, which is even lower on average than in the case without turnings. *CTG* and *Mix* do not suffer from gridlocks in this scenario.

**Travel time** - Fig. 20 (b) and (e): At short green times the travel time is for all policies higher than without turnings. This is explained by the fact, that without turnings, the vehicle need to stop once at each intersection. Vehicles turning in from cross traffic enlarge the platoons, so that the whole platoon cannot pass in one traffic light cycle. The travel time improvement of *CDG* and *Mix*, as well as the impact of the offset show similar characteristics to the case without turnings.

*4) Grid Scenario Simulation Results Summary*

From the simulation we learned the following about *CDG* applied on mutually influencing intersection in grids layout.

- In contrast to arterial scenarios, gridlocks may occur when *CDG* platoons longer than intersection interspaces arise.
- Gridlocks drastically reduce the benefit of *CDG* (in our scenario down to 15 %) with a simultaneous increased travel time.
- Countermeasures to avoid gridlocks are green times short enough to create platoons shorter than intersection interspaces.
- Offsets can mitigate gridlocks for one travelling direction.
- At very short green times, travelling times increase considerably, which is the case for all policies studied.
- Lower CDG penetration rates (Mix) are less vulnerable to gridlocks and of little potential for improvement.

## D. Vulnerability of CTG and CDG with respect to Gridlocks

From the grid layout simulation we observed that instability of the traffic flow increases with longer green times, high inflow rates (maximum possible in our case) and turn rates. This applies for *CDG* as well as for *CTG* and *Mix*. However, *CDG* is more sensitive in this regard. The high traffic density cause by the dense *CDG* platoons provides no buffer space like the *CTG* platoons which contract while slowing down. Hence, with *CTG* the traffic flow is stable for higher turn rates at maximum possible inflow rates than with *CDG*. In order to further explore this relationship, Fig. 21 compares the traffic flow stability of the grid scenario simulations. We distinguish between three states of stability which are determined from the metrics measured during simulation.

Fig. 19. Grid scenario simulation results without turnings

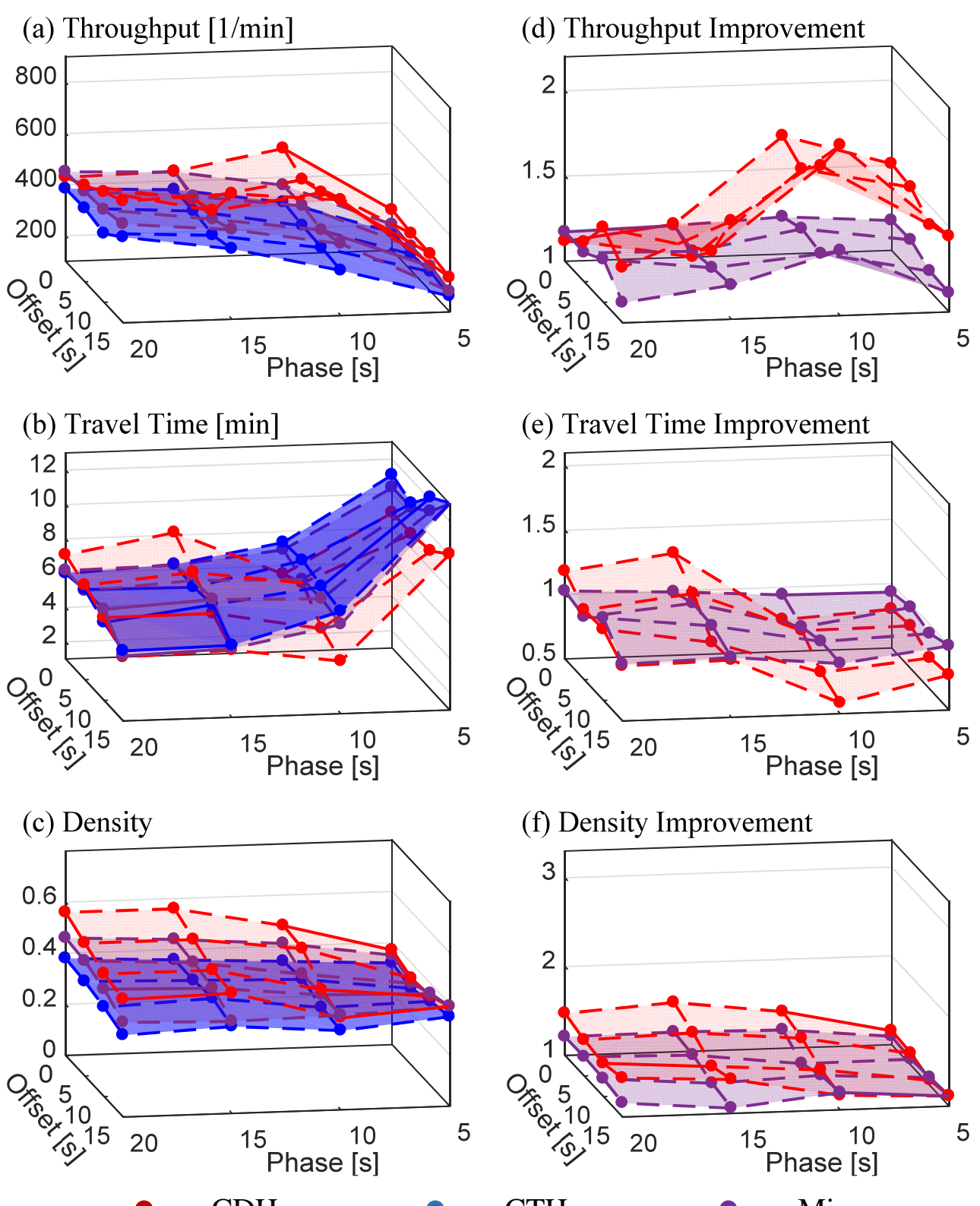

Fig. 20. Grid scenario simulation results with turnings

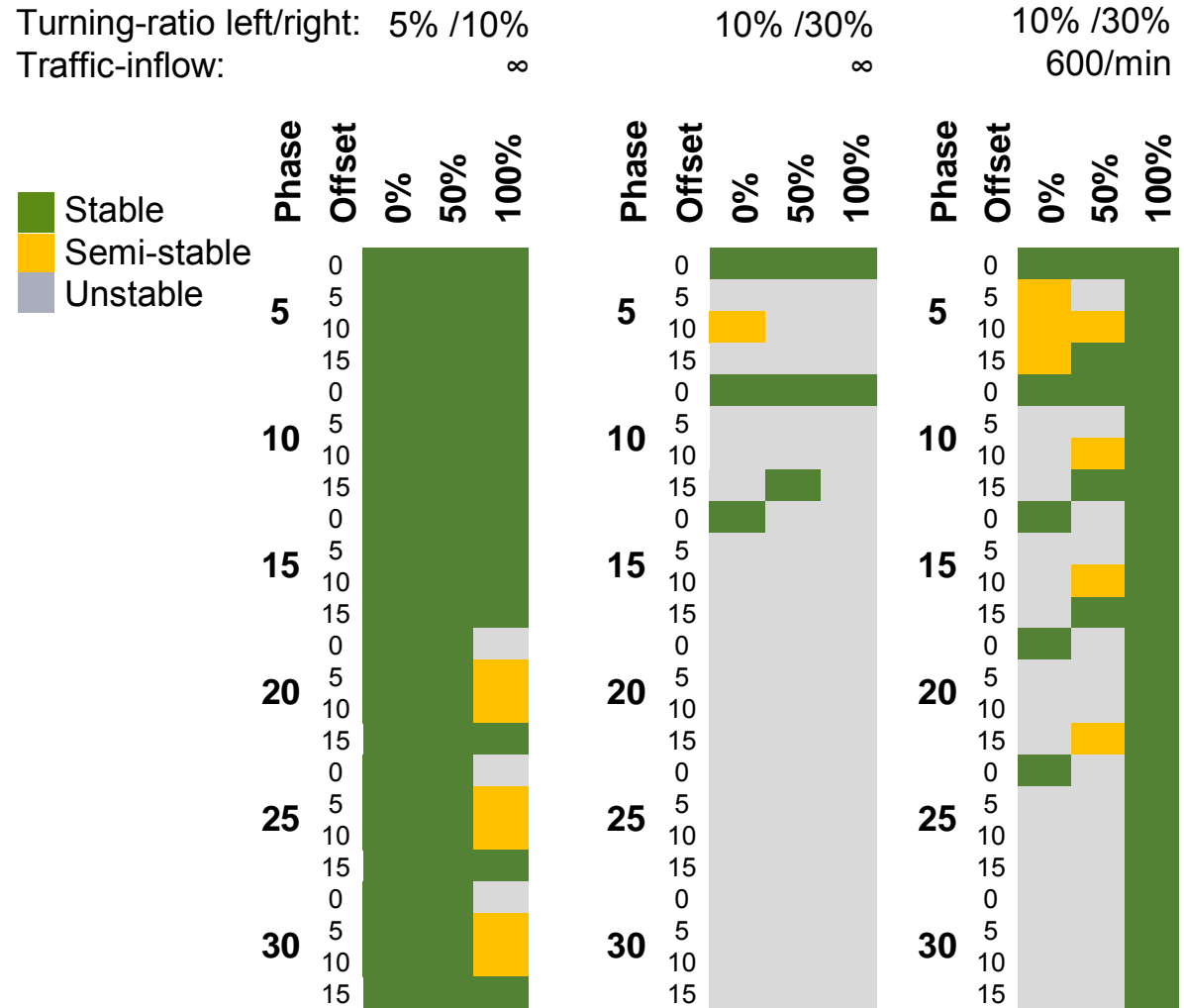

Fig. 21. Traffic flow stability in the grid scenario simulation

- Stable: all metrics increase to a steady state level at the beginning and keep that state for the whole simulation time (20,000 seconds). No gridlocks occur.
- Unstable: all metrics increase to a steady state level and at a certain point in time, the traffic flow collapses as e.g. depicted in Fig. 22. The throughput falls on a significant lower level, density and travel time increase, while the teleport [42] rate exceeds 1 %. This chart characteristic shows a completely jammed part in the middle of the grid due to gridlocks, while the outer intersections still have traffic throughput.
- Semi-stable: Gridlocks occur, however they can be dissolved so that metrics stay on a steady state level, while the teleport rate (see section VI.A) stays below 1 % of the throughput.

Fig. 21 shows that with maximum possible traffic inflow and 5 % / 10 % turnings *CDG* becomes unstable at 20 s green time when no offset is present, while *CTG* and *Mix* are stable. This correlates with the results depicted in Fig. 20. However, if the turning rate is raised to 10 % / 30 %, *CTG* also gets unstable. With the same turning rates and a lowered traffic inflow of 600 vehicles per hour, *CDG* gets even more stable than *CTG*. This is because of the higher throughput of *CDG*, which creates more free spaces than *CTG*. A further finding here is that offsets appear to have a negative influence on *CTG* stability as the only stable permutations are green times below 30 s without offset.

### *E. Conclusion Multi Intersection Performance*

Fig. 23 compares the throughput improvement of the grid scenario with single intersection and the arterial scenario with the single traffic light. Min and max refer to the offset with the best and the worst improvement respectively.

The performance of *CDG* in the grid without turnings is approximately the same as for a single traffic light up to 10 s green times. Above 10 s, the disturbance effects (see section VI.A) result in a considerable performance drop. A similar picture can be observed with turnings, with an additional performance drop when no offset is present. Additionally, the gridlock impact is higher with turnings at longer green times. The performance drop of *Mix* is quite less for all constellations. An exceptional case is 5 s green time where we have a very high performance at the single traffic light, due to discretization effects.

In the arterial scenario with 25 s to 35 s green time, disturbance effects are present for all *CDG* permutations. As *CTG* is not affected by them, we see an overall worse throughput improvement of *CDG* here. Additionally, *CTG* shows comparably high throughput in absolute numbers at long green times, which makes the impact of disturbance effects on the performance comparison with *CDG* grow. Thus, the arterial chart without turnings apparently is an extension of the grid chart. With turnings, the *CDG* benefit comes to the fore notably, as the flushing effect (see section VI.B) gets interrupted more often by the turning vehicles and *CDG* can reap the benefits of more start-ups similar to shorter green times.

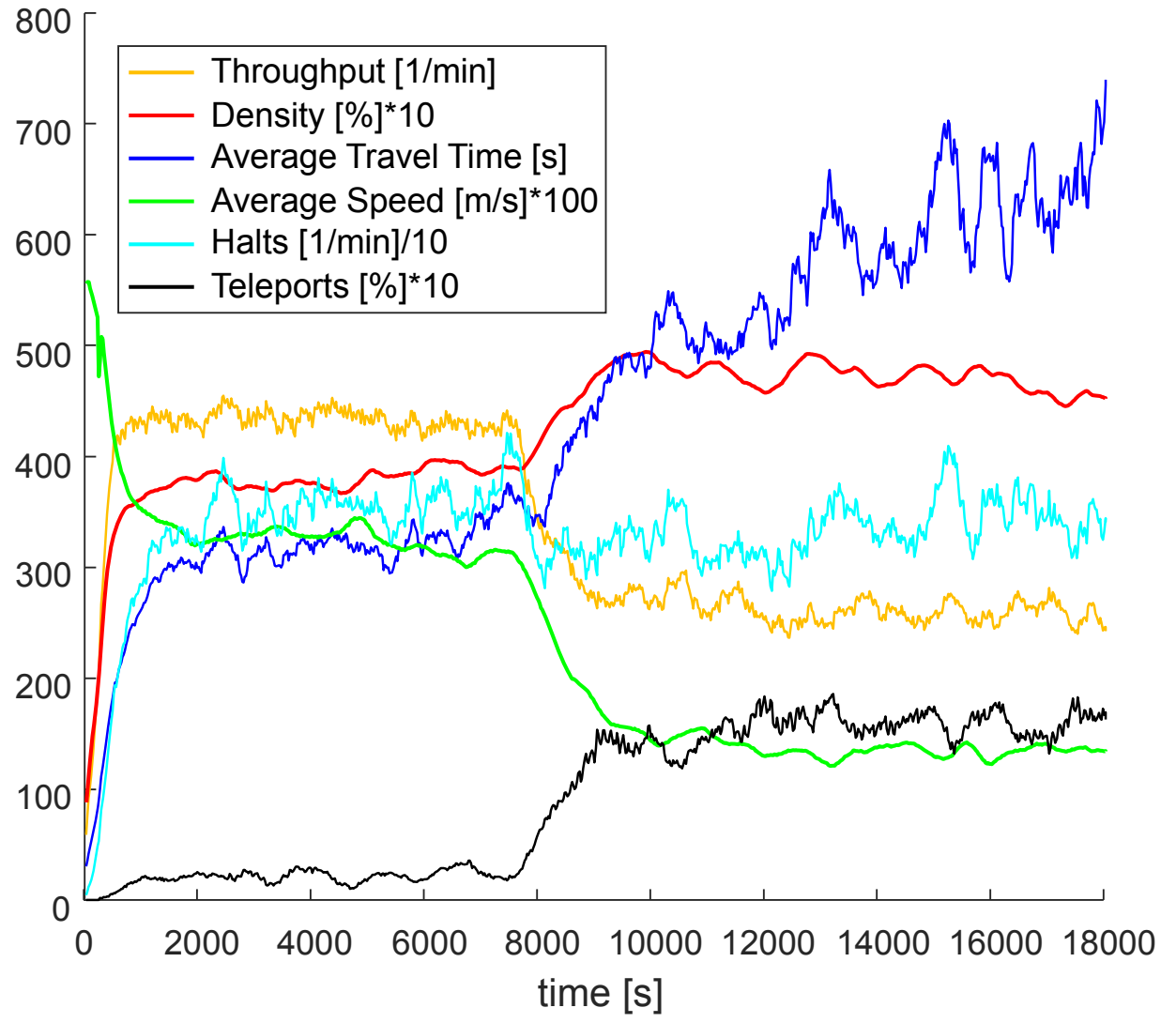

Fig. 22. Unstable simulation run

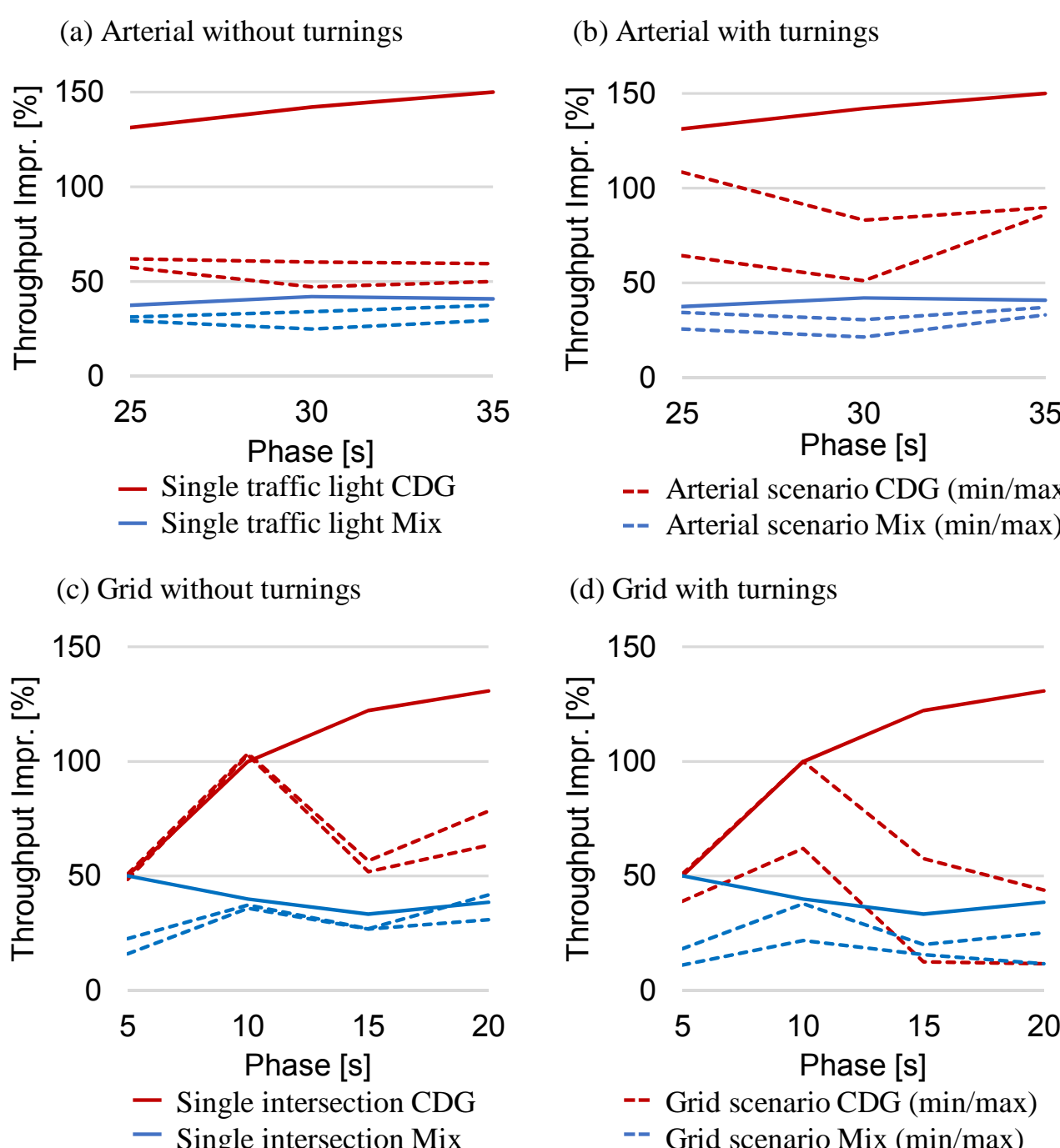

Fig. 23. Results of arterial and grid scenario simulation

In conclusion we found that *CDG* and *CTG* performance in multi intersection scenarios is influenced by many different effects. Their impact can be observed as a superposition in the measured metrics. In addition to the results presented in this work, the authors conducted further studies on each effect in order to explain them correctly. However, isolating each effect requires many more simulation scenarios, chart analytics and visual observation of simulations, which is beyond the scope of this work.

**Remark:** *For some cases in both scenarios, grid and arterial, we observed that an improved throughput of CDG over CTG, comes with less improvement of the travel time or even with a setback. Actually one would assume intuitively that travel time and throughput should be improved approximately proportionally. However, this is not the case when different inter vehicle distances are considered in signalized networks. If e.g. for situation 1 and 2 in section VI.B.1) the distances are halved, the throughput is approximately doubled, while traveling through the intersections takes the same time for each vehicle, except for a little less waiting time at the first queue.*

Our objective in this work was to present the overall benefit of one-vehicle look-ahead *CDG* in most common traffic scenarios. Our most relevant findings are summed up as follows.

- If the ratio of intersection interspaces and green time length is too high, *CDG* leads to disturbance effects in the traffic flow in the form of junction and turn blocking.
- In grid scenarios these disturbances provoke gridlocks when the traffic inflow is maximum possible, more likely than with CTG. For limited inflows, *CDG* is less sensitive for gridlocks than *CTG.*
- Offset positively counteracts such disturbance effects
- *CDG* penetration rates below 100 % are less sensitive to the disturbances. This improves the ratio between penetration rate and performance benefit of *CDG* considerably over single intersection scenarios. For some edge cases, a penetration of 50 % *CDG* even outperforms 100 % of *CDG* penetration.
- For all scenarios and parameter permutation tested, *CDG* improves traffic throughput. However for some situations, this improvement of *CDG* is bought by higher travel times due to its vulnerability to disturbance effects.

**Summary of Section VI**: *The performance of CDG in grids and arterial scenarios is sensitive to the traffic light configuration in relation to intersection interspaces. Green times above a certain threshold may lead to disturbance effects (junction blocking and turn blocking). An offset positively counteracts such disturbance effects. CDG showed an*

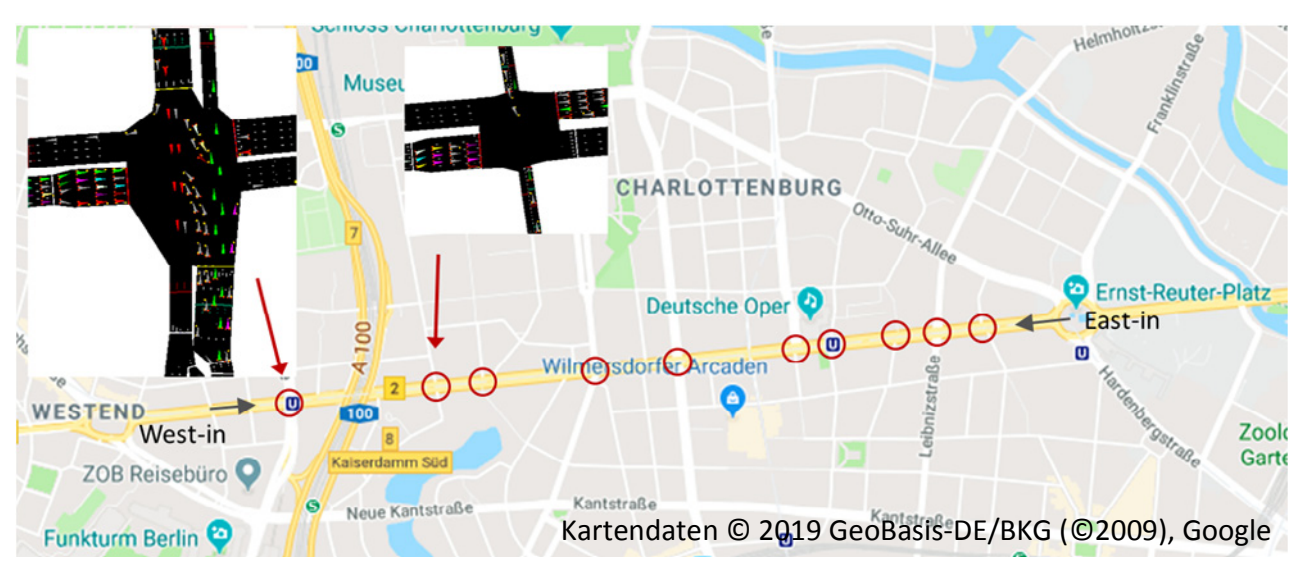

Fig. 24. Real world scenario: arterial road with nine intersections in Berlin

*improvement over the CTG baseline, in all cases. Finally, we emphasize that the discussed disturbance effects could be prevented by adding a cooperative aspect to CDG. If the vehicles in a platoon could anticipate an unintended stop within the intersection area, the general performance of CDG could be improved considerably.*

## VII. REAL WORLD ROAD NETWORK

In order to confirm the results gained in Section IV to VI using synthetic simulation scenarios, we now attempt to assess the real world performance of *CDG* in this section. For this purpose, we model a simulation scenario covering a heavily frequented arterial road in Berlin, Germany, as depicted in Fig. 24. This includes the Bismarckstraße between Theodor-Heuß-Platz and Ernst-Reuter-Platz with ten traffic light coordinated intersections with interspaces between 160 m and 500 m (266 m on average). The main difference to the synthetic scenarios in the previous section is the real world intersection layout, interspaces, and traffic light program including offset. While the road layout and the traffic light configuration is captured from real data, we again assume a maximum possible traffic inflow and an unobstructed outflow. Further assumptions without validation are the following.

- No pedestrians are blocking vehicles while turning.
- While assuming a capable cooperation concept to enable negotiation of lane changes between vehicles at high penetration rates of *CDG*, we excluded lane changing by respective route design in the previous sections. We now employ the SUMO lane changing model [38] without validating it analogous to Section V. As this model does not support opening gaps for merging parallel traffic, we accept a performance drop of *CDG*.
- Due to traffic backlog and quite large intersection interspaces, platoons of very large size appear, which in reality needs to be split to achieve platoon stability (see Section II.B for explanation). This splitting would slightly lower the performance of *CDG*.

### *A. Simulation Setup*

The traffic light program was observed on week-days between 10 am and 12 am. Public authorities indicated a fixed schedule for this period (dynamic priority phases e.g. for buses neglected). Table 3 lists the phase times of the program for each intersection in the following order: 1) green on major road, 2) yellow, 3) clearance interval, 4) protected left turning major road, 5) green on minor road, 6) yellow, 7) clearance interval.

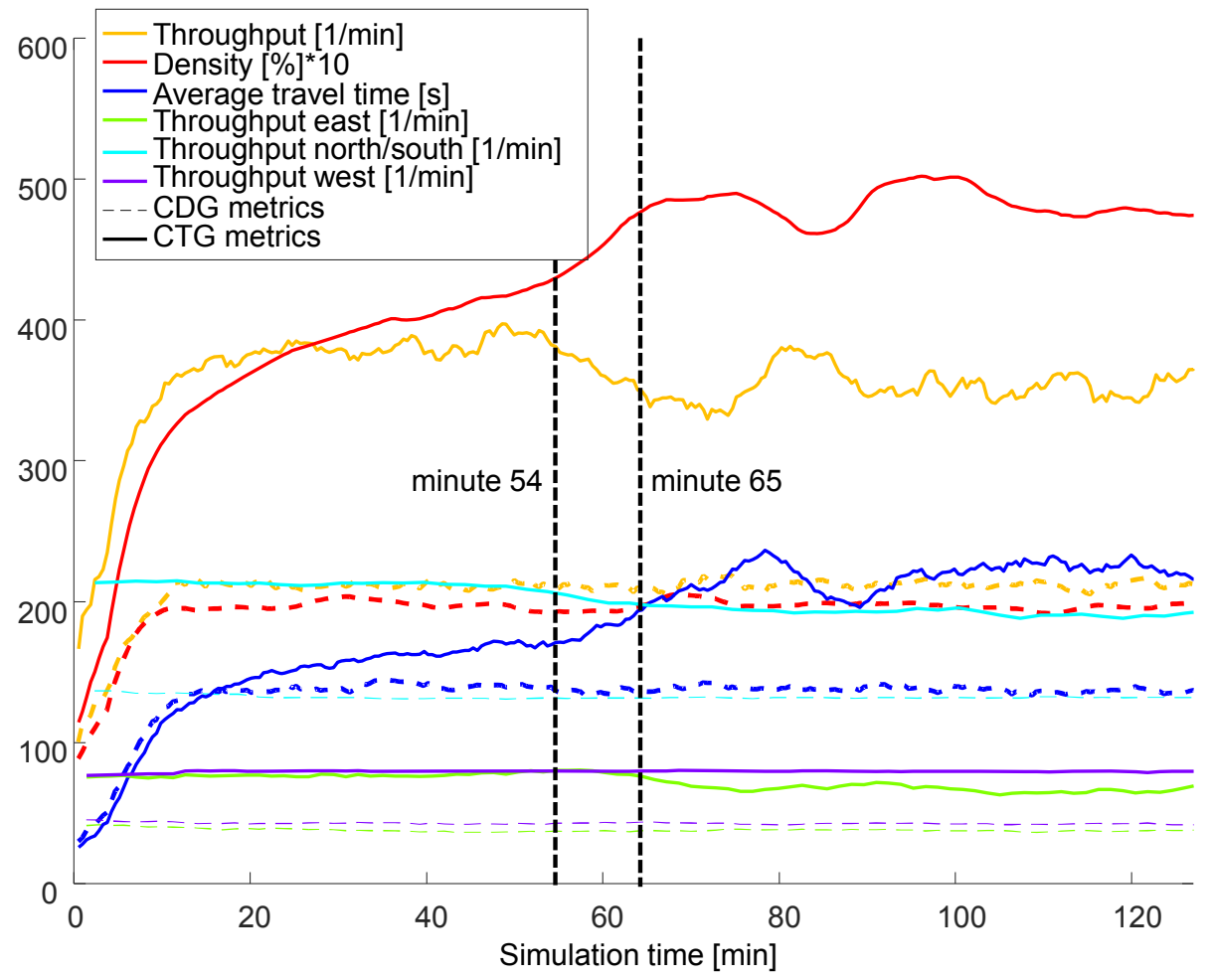

Fig. 26. Simulation results real world scenario Berlin CDG / CTH

The base ratio for turning was estimated by observation at 80 %, 12 %, 8 % (straight, right, left) on average on the major roads and 75 % / 16 % / 9 % on the minor roads. The final turning configuration was adjusted based on the number of lanes per direction at each intersection, as listed in Table 3. Combining the real world traffic light program with this setup leads to a simulation setup using SUMO's default driver model, with all lanes evenly occupied and without traffic jams.

## B. *Evaluation*

Fig. 26 compares the results of two simulation runs with *CDG* and *CTG* for the first two hours simulation time. In addition to the metrics used before, the figure also separately indicates the ingoing traffic flow from east, west, and from the minor roads (north and south). *CTG* reaches a steady state level for all metrics after 15 minutes simulation time, with a throughput of around 210 vehicles per minute. *CDG* reaches around 380 vehicle per minute, however the density and the travel time keep rising slightly. After minute 65 the metrics begin to stabilize while the throughput drops slightly to around 355 vehicle per minute. This behavior is the result of an east bound traffic backlog at Suarezstr. The traffic light there shows a slightly lower capacity than Kaiser-Friedrich-Str. in the simulation scenario. As *CDG* leverages the longer green times better than *CTG*, assuming a maximum possible traffic inflow, this leads to a larger capacity difference and, thus, to a rising backlog. The backlog reaches Am Schillertheater at minute 54 and finally reaches the east traffic inflow at minute 65. This becomes apparent with the declined inflow rate east. Once the traffic jam emerges, vehicles have difficulties finding gaps for lane changes, due to the close vehicle interspaces of *CDG* and missing cooperative lane change maneuvers. Thus, some vehicle reach the intersection in the wrong lane and block that lane for a whole cycle. This further reduces the intersection capacity and the traffic jam cannot be dissolved. However, even with the named drawbacks, in terms of throughput, *CDG* still outperforms *CTG*. Applying, a switch to *CTG* at 30 Km/h solves this problem completely. As shown in Fig. 25, *SWITCH1* reaches the same throughput on average as *CDG*.

TABLE 3
BERLIN SIMULATION SCENARIO CONFIGURATION

| Minor street name | Traffic light program | Offset, Distance | Turnings major | Turnings minor |
|---|---|---|---|---|
| Königin-Elisabeth-Str. | 22/3/8/3/11/3/10 | 54, 500 | .8/.12/.08 | .4/.3/.3 |
| Sophie-Charlotte-Str. | 24/3/9/0/12/3/9 | 19, 160 | .8/.12/.08 | .5/.3/.2 |
| Witzlebenplatz | 27/3/9/0/9/3/9 | 31, 390 | .88/.12/0 | 0/1/0 |
| Suarezstr | 24/3/9//012/3/9 | 55, 290 | .8/.12/.08 | .75/.16/.09 |
| Kaiser-Friedrich-Str. | 28/3/9/0/8/3/9 | 32, 250 | .8/.12/.08 | .75/.13/.12 |
| Wilmersdorfer Str. | 23/3/9/0/13/3/9 | 46, 280 | .8/.12/.08 | .75/.16/.09 |
| Krumme Str. | 24/3/9/0/12/3/9 | 31, 160 | .8/.12/.08 | .75/.16/.09 |
| Pedestrian Lights | 29/3/9/0/7/3/9 | 42 (31), 210 | 1/0/0 | 0/0/0 |
| Leibnitzstr. | 22/3/9/0/14/3/9 | 5, 160 | .8/.12/.08 | .5/.3/.2 |
| Am Schillertheater | 27/3/9/0/9/3/9 | 0, 0 | .88/.12/0 | 0/1/0 |

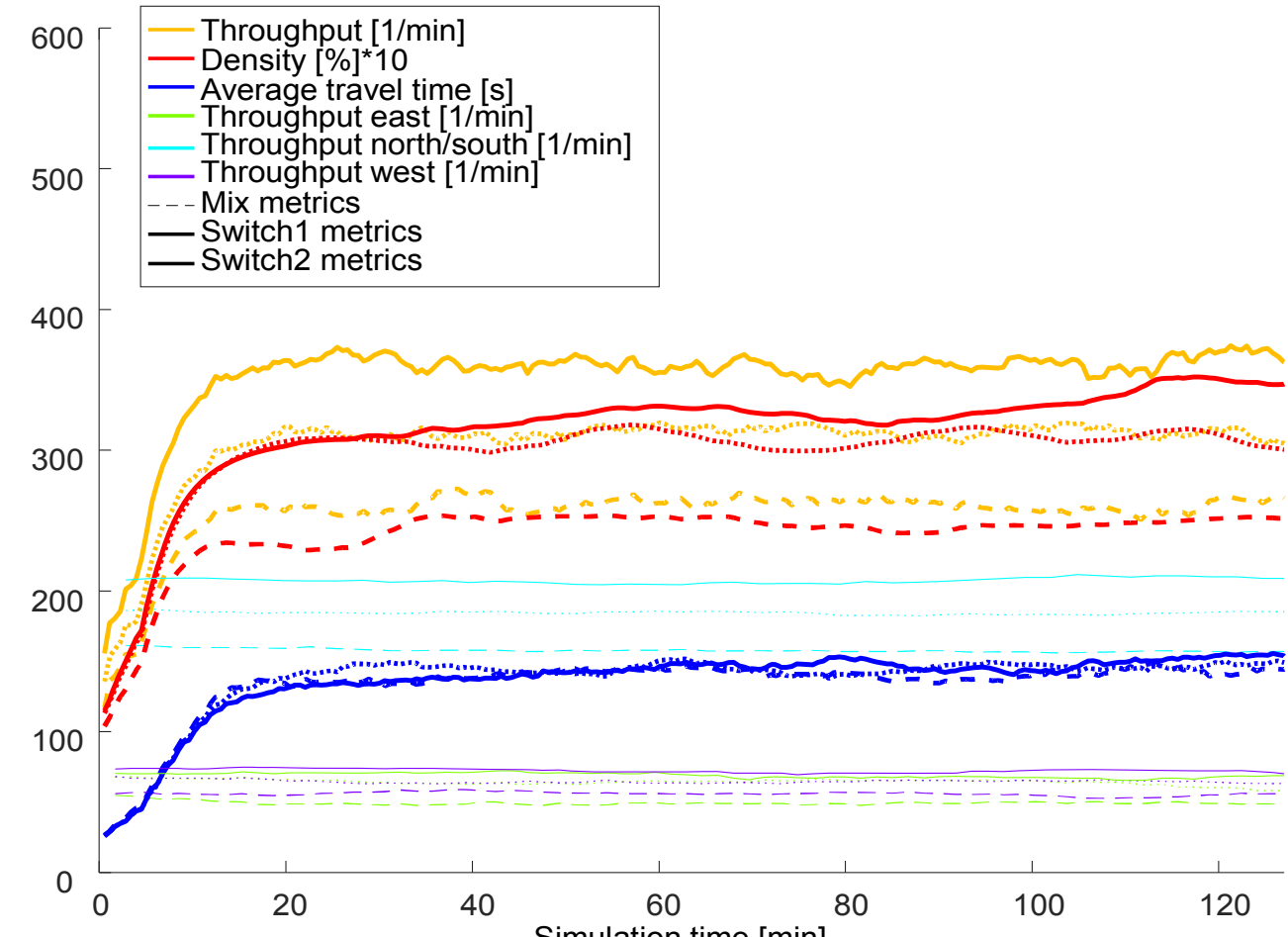

Fig. 25. Simulation results real world scenario Berlin Mix / Switch1 / Switch2

## C. *Conclusion*

In the previous section, we used synthetic simulation scenarios to reveal the relationship of different constellations between road topology and traffic light configuration. This real world road network scenario in contrary shows the performance of *CDG* in a real world traffic system including a plethora of such constellations at the same time. Moreover, in the previous section we neglected the impact of lane changes by route design, as we assume a cooperative merging feature coming with 100 % penetration of *CDG*. In this section we included uncoordinated lane changing which led to a jammed condition for *CDG*. However, we showed that this effect would not necessarily occur in real world, as it is due to the non-cooperative character of merging in the simulation models of SUMO combined with small gaps. Besides that, even with a big part of the scenario in a jammed condition, *CDG* still outperforms *CTG* in terms of traffic throughput. While the travel time raises by 60 %, the *CDG* throughput is 70 % higher than *CTG*. The following consideration pertain to the performance of *CDG* before the jamming occurred. Regarding throughput improvement of *CDG*, the real world road network scenario matches the results of the arterial scenario in Section VI.B for the configuration of 25 s green time and no offset. The throughput improvement of *MIX* is slightly lower. We observe no negative impact by the presence of offset and no considerable disturbance effects (see section VI.A) before minute 65. This becomes apparent in particular by the steady inflow from the minor roads. The absence of disturbance effects is a result of the very well balancing of traffic light configuration to the intersection interspaces done by the Berlin traffic management. Given the assumption that we usually find such well balancing in traffic management, *CDG* can exploit much of its potential in traffic systems, not only at single intersections. Surprisingly, the travel time is almost equal for all policies. Even with visually observing the simulation, we could not find a clear cause. The most reasonable explanation here is the following. As we learned from the previous section that *CDG* can buy throughput by travel time, the specific configuration of the scenario might lead to levelling out the travel time by different throughputs and densities for each policy.

**Summary of Section VII**: *In a real world network with a well-balanced traffic light configuration, CDG can exploit much of its potential in traffic systems, not only at single intersections. Comparing the performance of SWITCH1 and CDG in this scenario and in Section III, we could deduce the following finding. In dense, urban traffic systems a switch from CDG to CTG at 30km/h is recommended in order to create gaps for lane changes. At single intersections, for example, on crossing rural roads, this is not required and CDG without switching results with a considerably better performance.*

## VIII. CONCLUSION & FUTURE WORK

In this paper, we comprehensively investigated the impact of applying a constant distance gap (CDG) policy for starting platoons at traffic lights. The applicability of CDG in real traffic is limited, due to its demand on complex communication topologies in order to achieve string stability. However, we were able to show its capability to increase the capacity of traffic light controlled intersections.

As a baseline for comparison, we calibrated a constant distance gap (CDG) policy in the vehicle dynamics simulation PHABMACS using real word driving data. Compared with this baseline, CDG increased the capacity of a single intersection by up to 140%, depending on the green light time and the ratio of turning vehicles. The penetration rate of CDG in mixtures with CTG does not have a linear impact on the capacity enhancement on single intersections, which is a clear downside. A penetration of 50% still peaked with a capacity enhancement of 40%.

For large scale analysis of CDG performance on multiple adjacent intersections in traffic systems, we employed traffic simulation with several thousand vehicles. To achieve this scaling, we proposed a method for calibrating and validating traffic simulation against vehicle dynamics simulation. This calibration enables traffic simulation to render the same results as vehicle dynamics simulation regarding the relevant metrics.

The large scale analysis yielded the following conclusions:

- Compared to single intersections, a full penetration of CDG reaches a lower performance at arterial roads and grids with multiple intersections due to occurring disturbance effects. This performance drop is less pronounced at lower CDG penetration rates.
- CDG outperformed CTG regarding throughput in all cases observed in this work. Although, a 50 % penetration rate of CDG has less potential for improvement, it is less vulnerable to disturbance effects and appears as stable as CTG in traffic systems.
- While CDG is more prone to gridlocks in traffic grids at maximum traffic inflow, it is less prone to gridlocks than CTG if the inflow is limited.
- CDG gains a considerable travel time improvement on arterial roads. However, the increased throughput of CDG comes with a higher density in traffic grids, which may lead to an increased average travel time.

After exposing the edge cases using synthetic scenarios with uniform parameterization, we finally modeled a real world road network scenario which includes a mixture of parameterizations. This mixture originates from the heterogeneous road geometry in Berlin, Germany and its well calibrated traffic light configuration. CDG improved the traffic throughput by 80% at the same average travel time as CTG. Given the average green light time and turning rates, this improvement confirms the results of a single intersection.

The simulation results revealed a potential performance drop of CDG originating from prevented lane changing and blocked intersections due to missing coordination and small gaps. Both problems could be tackled by a close range coordination between vehicles [37], to create gaps for merging and prevent entering intersections when a stop within the intersection area is likely. Given such coordination, the potential of performance improvement for CDG in a traffic system seems similar to the single intersections.

Our future work includes implementing a coordination strategy as described above and a real world road network scenario for traffic grids. Replacing the maximum traffic inflow by real world traffic flows at rush hours will reveal information about the benefit of CDG by a market introduction in today's traffic.

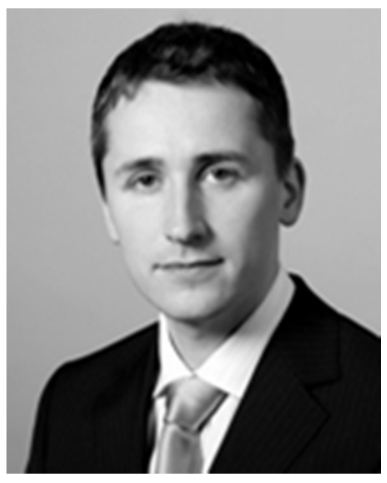

**Kay Massow** received the Diploma in computer engineering from the Technical University of Berlin, Berlin, Germany, in 2008. In the past, he has worked for Daimler and Volkswagen of America. He is now team leader at the Department of Automotive Services and Communication Technologies, Fraunhofer Institute for Open Communication Systems, Berlin. Additionally, he assists teaching and research at the Daimler Center for Automotive IT Innovations, a joint initiative of the Daimler AG and the Technical University of Berlin. He is currently working in the fields of intelligent transport systems, cooperative driving applications, digital high definition maps, and automotive big data analytics.

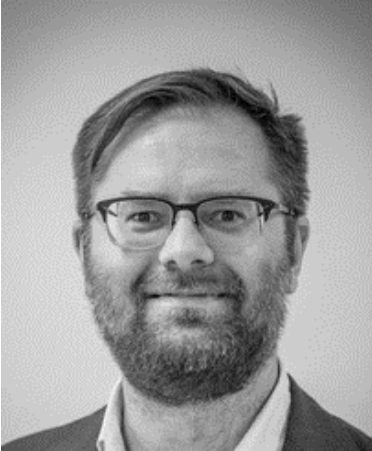

**Ilja Radusch** received the Ph.D. degree in engineering from Technical University of Berlin, Berlin, Germany. He is currently the Head of the Department for Automotive Services and Communication Technologies, Fraunhofer Institute for Open Communication Systems, Berlin, and the Managing Director of the Daimler Center for Automotive Information Technology Innovations, Technical University of Berlin. His research and teaching interests include (secure) car-to-X communications, Internet-based telematics services, and simulation for cooperative vehicles.

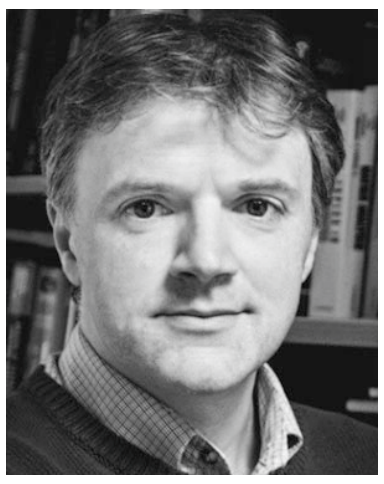

**Robert Shorten** (SM’17) received the B.E. degree in electronic engineering and Ph.D. degree from the University College Dublin, Dublin, Ireland, in 1990 and 1996, respectively. From 1993 to 1996, he was the holder of a Marie Curie Fellowship and was with the Daimler-Benz Research, Berlin, Germany, to conduct research in the area of smart gear-box systems. Following a brief spell with the Center for Systems Science, Yale University, working with Prof. K. S. Narendra, he returned to Ireland as the holder of a European Presidency Fellowship in 1997. He is a cofounder of Hamilton Institute, National University of Ireland, Maynooth, Ireland, where he was a Full Professor until March 2013. He was also a Visiting Professor with the Technical University of Berlin from 2011 to 2012. From 2013 to 2015, he led the Control and Optimization team at IBM Research, Dublin. He was Professor of Control engineering and Decision science with University College Dublin from 2015-2019 as well as with IBM Research, and is now Professor of Cyber-Physical Systems Design at Imperial College London. His research spans a number of areas. He has been active in computer networking, automotive research, collaborative mobility (including smart transportation and electric vehicles), as well as basic control theory and linear algebra. His main field of theoretical research has been the study of hybrid dynamical systems.